Featured Article

# Neural correlates of episodic memory in the Memento cohort

Stephane Epelbaum[a,b,*,1], Vincent Bouteloup[c,d,1], Jean F. Mangin[e,f], Valentina La Corte[g,h], Raffaela Migliaccio[a,b], Hugo Bertin[e,i,j], Marie O. Habert[e,i,j], Clara Fischer[e,f], Chabha Azouani[e], Ludovic Fillon[e], Marie Chupin[e], Bruno Vellas[k,l], Florence Pasquier[m], Jean F. Dartigues[c,n], Fréderic Blanc[o], Audrey Gabelle[p,q], Mathieu Ceccaldi[r,s], Pierre Krolak-Salmon[t], Jacques Hugon[u], Olivier Hanon[v], Olivier Rouaud[w], Renaud David[x], Genevieve Chêne[c,d], Bruno Dubois[a,b], Carole Dufouil[c,d], for the Memento Study group**

[a]Institut de la mémoire et de la maladie d'Alzheimer, Département de neurologie, Hôpital de la Pitié Salpêtrière, Paris, France
[b]Sorbonne Universités, UPMC Univ Paris 06 UMR S 1127, and Inserm, U 1127, and CNRS UMR 7225, and ICM, Paris, France
[c]Inserm, Bordeaux Population Health Research Center, UMR 1219, University Bordeaux, ISPED, CIC 1401-EC, Univ Bordeaux, Bordeaux, France
[d]CHU de Bordeaux, Pole Santé Publique, Bordeaux, France
[e]CATI Multicenter Neuroimaging platform, http://cati-neuroimaging.com, Gif sur Yvette, France
[f]Neurospin, DRF, CEA, Paris Saclay University, Gif sur Yvette, France
[g]Institut de psychologie, Université Paris Descartes, Sorbonne Paris Cité, Paris, France
[h]Inserm UMR 894, Centre de psychiatrie et neurosciences, Laboratoire mémoire et cognition, Paris, France
[i]Sorbonne Universités, UPMC Univ Paris 06, CNRS, INSERM, Laboratoire d'Imagerie Biomédicale, Paris, France
[j]Département de Médecine Nucléaire, Hôpital de la Pitié-Salpêtrière, AP-HP, Paris, France
[k]Gérontopôle, Centre Hospitalier Universitaire de Toulouse, Toulouse, France
[l]Inserm UMR1027, Université de Toulouse III Paul Sabatier, Toulouse, France
[m]Univ Lille, Inserm 1171, CHU, Centre Mémoire (CMRR) Distalz, Lille, France
[n]CHU de Bordeaux, CMRR, Pôle Neurosciences, Bordeaux, France
[o]Centre Mémoire (CMRR), Hôpital Universitaire de Strasbourg, Département de Gériatrie, Hôpital de Jour Gériatrique, Strasbourg, France
[p]Department of Neurology and Memory Research and Resources Center, Gui de Chauliac University Hospital, Montpellier, France
[q]CHRU de Montpellier, Université de Montpellier, Institut of Regenerative Medicine and Bio-therapy (IRMB), INSERM U1183, CCBHM, Laboratoire de Biochimie Protéomique Clinique, Montpellier, France
[r]Université Aix-Marseille, INSERM, Institut des Neurosciences des Systèmes (INS) UMR 1106, Marseille, France
[s]APHM, Hôpitaux de la Timone, Service de Neurologie et de Neuropsychologie, Marseille, France
[t]Clinical and Research Memory Center of Lyon, Hôpital des Charpennes, Hospices Civils de Lyon, Lyon, France
[u]Centre Mémoire (CMRR) Paris Nord Ile de France, Groupe Hospitalier Lariboisiere FW Saint-Louis, APHP, Université Paris Diderot, Paris, France
[v]Service de Gériatrie, Université Paris Descartes, Hôpital Broca, Paris, France
[w]Département de Neurologie, Hôpital universitaire et faculté of Médecine, Dijon, France
[x]Centre Mémoire de Ressources et de Recherche, CHU de Nice, EA COBTeK, Université Côte d'Azur, Nice, France

**Abstract**

**Introduction:** The free and cued selective reminding test is used to identify memory deficits in mild cognitive impairment and demented patients. It allows assessing three processes: encoding, storage, and recollection of verbal episodic memory.

**Methods:** We investigated the neural correlates of these three memory processes in a large cohort study. The Memento cohort enrolled 2323 outpatients presenting either with subjective cognitive decline or mild cognitive impairment who underwent cognitive, structural MRI and, for a subset, fluorodeoxyglucose–positron emission tomography evaluations.

**Results:** Encoding was associated with a network including parietal and temporal cortices; storage was mainly associated with entorhinal and parahippocampal regions, bilaterally; retrieval was associated with a widespread network encompassing frontal regions.

[1]Both authors contributed equally to the article.
**The Memento Study group is described in the online Supplementary data section.

*Corresponding author. Tel.: +33142167525; Fax: +33142167504.
E-mail address: stephane.epelbaum@aphp.fr

**Discussion:** The neural correlates of episodic memory processes can be assessed in large and standardized cohorts of patients at risk for Alzheimer's disease. Their relation to pathophysiological markers of Alzheimer's disease remains to be studied.

## 1. Introduction

Episodic memory refers to memory for personal experience with respect to time and context [1]. The three principal processes involved in episodic memory are encoding, storage, and retrieval of information. It can be impaired in various diseases, for example, depression [2], Parkinson disease [3], frontotemporal dementia [4]. Alzheimer's disease (AD) is the most frequent disorder characterized by memory impairment [5]. Indeed, impairments in episodic memory performance are considered as the first clinical sign of typical AD and have been associated with atrophy of the entorhinal cortex and hippocampus [6–8].

A few longitudinal studies of cognition in healthy older adults have shown that a subtle decline in episodic memory often occurs before the emergence of the functional and overt cognitive changes required for a clinical diagnosis of AD dementia [9–14]. These findings led to the amnestic mild cognitive impairment (MCI) [15] concept, a predementia condition in elderly individuals, which is characterized by subjective and objective memory impairments with relatively preserved general cognition and functional abilities. Before this MCI stage of AD, subjective cognitive decline (SCD) can be a symptom of preclinical AD [16,17].

The Free and Cued Selective Reminding Test (FCSRT) has been proposed as a verbal associative episodic memory test [18]. It aims at exploring the three memory processes in a single neuropsychological test and is used in clinical practice of some memory clinics for AD diagnosis [11,19]. Its subscores allow the assessment of serial cognitive processes involved in episodic memory: immediate recall (IR) for encoding, index of sensitivity to cueing (ISC) for storage, and total free recall (FR) for retrieval of memorized stimuli. Previous studies have shown that FCSRT is useful for prognosing MCI patients who will decline to dementia stage of AD [11,20–22] and for diagnosing typical amnestic AD patients among various neurodegenerative conditions [19]. When storage is impaired (i.e., low FR score and low total recall or ISC scores), an amnestic syndrome [23] termed "of the hippocampal (or medial temporal) type" has been defined. However, only a few imaging studies, mainly in a small number of demented patients, have shown a correlation between hippocampal volumes and FCSRT performances [24,25]. In addition, little is known on the link between FCSRT performances and other brain regions known to be implicated in episodic memory such as the working memory network [26] or prefrontal areas [27]. There is a large number of studies that tackled the question of the neural correlates of episodic memory (for a recent review see [28]). However, the experimental paradigms frequently differ from one study to the next, which induced some discrepancies, concerning for instance the laterality of the medial temporal lobe involvement found to be mainly left sided in some studies [29–31] right sided in other [32,33] and sometimes bilateral [34,35].

In this study, we investigated structural and metabolic correlates of the three episodic memory processes assessed by the FCSRT in a large French cohort of participants with standardized cognitive assessment as well as structural and metabolic imaging. Within this framework, we hypothesize that encoding and storage phases would be related to hippocampal and parietal regions, and recollection phase would be related to a widespread brain network, including more anterior brain regions. Our large sample size and standardization allow us to draw unequivocal conclusions from our results, shedding some light on previously described discrepancies [36].

## 2. Material and methods

### 2.1. Participants

Memento study consecutively enrolled 2323 nondemented outpatients in 28 French expert memory clinics, from 2011 to 2014. The study procedures and participants' baseline characteristics are described elsewhere [37]. At inclusion, participants presented either with cognitive impairment, when performing worse than one standard deviation to the mean of a group (with similar age, age/educational norms) in one or more cognitive domains, this deviation being identified for the first time through cognitive tests performed recently (less than 6 months preceding screening phase), or with isolated cognitive complaints, if participants had subjective cognitive complaint (assessed through visual analogic scale), without any objective cognitive deficit as defined previously, while being 60 years and older, and they all had a Clinical Dementia Rating scale [38] score ≤0.5. Main exclusion criteria were contraindication or refusal to perform magnetic resonance imaging (MRI), neurological disease such as treated epilepsy, treated Parkinson's disease, Huntington disease, or brain tumor, history of head trauma with neurological sequelae, stroke occurring in the past three months, history of schizophrenia, or illiteracy.

All examinations (including neuropsychological battery administration, clinical examinations, brain MRI, and fluorodeoxyglucose [FDG] positron emission tomography [PET]) performed through Memento followed standardized procedures.

The analytic sample consists in participants who underwent a brain MRI and a neuropsychological evaluation, including the FCSRT at their inclusion in the cohort (N = 2157). A subsample that additionally performed the optional FDG-PET was considered in a subsequent analysis (N = 1310).

All participants signed an informed consent to participate in the study that was approved by the ethics committee "Comité de Protection des Personnes Sud-Ouest et Outre Mer III." The study was conducted following standards of the Good Clinical Practice and the Helsinki Declaration. Although not a clinical trial, the protocol was registered in ClinicalTrials.gov (Identifier: NCT01926249, https://clinicaltrials.gov/ct2/show/NCT01926249).

### 2.2. Neuropsychological evaluation

A full neuropsychological test battery was administered to participants at baseline [37] including the FCSRT [39] to study the verbal episodic memory. In this associative memory test, the subject has to learn 16 words by groups of four with each corresponding cue provided verbally by the tester (e.g., "fish" is the cue for the word "herring"). In a first step, the subject is asked to recall words just after reading them, four by four (namely IR, scored from 0 to 16). Then, three recall (firstly free and then cued) trials separated from each other by a distractive task (mental calculation during 20 seconds) are successively performed. The FR score ranges from 0 to $16 \times 3 = 48$. The ISC is computed as $100 * (\text{sum of the three cued recall}/[48\text{-FR}])$. The list of the 16 words and the detailed procedure of execution are available elsewhere [40].

The neuropsychological test battery also included the Rey figure copy [41] that assesses visuospatial and visuoconstructive abilities and was used as a control of the specificity of the morpho-metabolic correlates of the FCSRT subscores.

Using performances at the full neuropsychological tests battery, Petersen criteria [42] were applied to categorize participants' cognitive status as non-MCI (SCD), pure amnestic MCI, multidomain amnestic MCI, pure nonamnestic MCI, multidomain nonamnestic MCI.

### 2.3. MRI evaluation

Brain magnetic resonance images were acquired after a standardization of the imaging processes (notably the sequences used) by a dedicated neuroimaging specialist team (CATI for "Centre pour l'Acquisition et le Traitement des Images", http://cati-neuroimaging.com/). MRI machines of 1.5 and 3 Tesla were used for this study (the complete list of machines is provided in Supplementary Appendix A).

All MRI scans were centralized, quality checked, and postprocessed by the CATI to obtain standardized measurements for each participant. The MRI protocol included 3D-T1 1 mm isometric sequences that were used to assess the whole-brain, gray matter, and white matter volumes with Statistical Parametric Mapping [43], hippocampal volumes with the SACHA software [44,45] and cortical thickness with FreeSurfer in Desikan-Killiany atlas [46,47].

### 2.4. FDG-PET evaluation

As for MRI, the CATI allowed for intercenter reproducibility through harmonization of FDG-PET protocols and postprocessing [48]. Structural MRI images were coregistered to PET images using Statistical Parametric Mapping 8 with visual inspection to detect any coregistration errors. MRI 3D T1-weighted images were segmented and spatially normalized into the Montreal Neurological Institute space using the VBM8 package (http://dbm.neuro.uni-jena.de/vbm/) implemented in Statistical Parametric Mapping 8. MRI matrix transformation was then used to spatially normalize PET images into Montreal Neurological Institute space. Parametric PET images were created for each individual, by dividing each voxel with the mean activity extracted from the reference region, the pons. Finally, gray matter masks extracted from each individual MRI volume were applied to the parametric PET images before Regions Of Interest (ROIs) analysis. Metabolic FDG-PET indexes were calculated in ROIs from the Automated Anatomical Labeling 2 (AAL2) atlas [49] to the exception of the cerebellum.

### 2.5. APOE genotyping

Apolipoprotein E (*APOE*) ε2, ε3, or ε4 alleles were determined for all participants by KBiosciences (Hoddesdon, UK; www.kbioscience.co.uk) as described elsewhere [37].

### 2.6. Statistical methods

Sample characteristics are reported as median (q1; q3) or frequency, as appropriate. Between-group comparisons were performed through the $\chi^2$ test for discrete variables or analysis of variance tests for continuous variables. Multivariable analyses were undertaken on three outcomes (IR, FR, and ISC scores at inclusion). As more than half of the population scored 16 (maximum score) in the IR subscore, it was dichotomized as equal to 16 versus <16, and logistic regressions were computed for analyses. To account for skewed distributions of FR and ISC subscores, their anatomical and metabolic correlates were modeled through median regression. For each outcome, models were built using brain structure as the "exposure" of interest and gender, age, education, number of ε4 alleles of *APOE* genotype, and type of MRI/PET as adjustment covariates. Due to multiple comparisons in 34 cortical thicknesses (FreeSurfer) MRI ROIs and in the 47 (AAL2) FDG-PET ROIs, a false discovery rate (FDR) was maintained at 0.05 or less by recomputing *P*-values [50]. A *P*-value ≤.05 was considered statistically significant. The effect sizes (estimate) of the significant association were then presented graphically on an inflated brain mesh.

Finally, we analyzed jointly the association between the imaging measurements and the three FCSRT subscores, assuming that all three are markers of the episodic memory. As the episodic memory in itself is unmeasured, we used a latent class analysis approach [51]. This method allows linking multiple outcomes of different nature (i.e., binary, ordinal, discrete and continuous) generated by the same underlying latent process and flexible enough to deal with nonlinear associations. We thus can estimate whether imaging measurements are associated with the latent process and, using contrasts, test whether the contribution of the subscores can be considered statistically equivalent or different [52].

Analyses were performed using SAS software version 9.3 (SAS Institute, Cary, NC) and R (LCMM package v1.7.8) for latent class analysis.

Finally, to determine whether *APOE* ε4 genotype could modify the relation between cortical thickness and FCSRT performances, we introduce *APOE* ε4 genotype*cortical thickness interaction term and was tested in the models. Uncorrected and FDR-corrected *P*-values were computed. For uncorrected *P*-values < .05, results of stratified analyses (Non–*APOE* ε4 and *APOE* ε4 carriers) were presented.

## 3. Results

Of the 2323 participants, 2157 (age median and [interquartile range]: 71.6, [65.6–77.1] years) were administered the FCSRT and had a brain MRI. Among them, 1310 underwent FDG-PET scan (Fig. 1). Table 1 shows the analytical sample baseline characteristics according to FDG data availability. Participants who had a FDG-PET were more likely women (67% vs. 59%, *P* = .0002), had less frequently a clinical dementia rating score equal to 0.5 (63% vs. 58%, *P* = .018) and had more frequently an SCD or a nonamnestic MCI profile (*P* = .045). The scatter plots of raw FCSRT subscores are provided in Supplementary Appendix B both globally (whole cohort) and by *APOE* and cognitive (SCD or MCI) status.

### 3.1. MRI measures and FCSRT subscores

Fig. 2 summarizes results of MRI analyses and FCSRT scores correlations. As expected, greater hippocampal volume was associated with increased odds of having high scores at IR: 1.30 (1.15; 1.46) (odds ratio [<16 vs. = 16] [95% confidence interval {CI}]), FR and ISC: 5.53 (4.72; 6.35) and 3.72 (2.71; 4.73), respectively (differences in median [95% CI]), all FDR corrected *P* values < .0001 associations. Distinct patterns of associations were observed for the three subscores: FCSRT-IR and FCSRT-ISC were mainly associated with entorhinal and parahippocampal regions, bilaterally; FCSRT-FR was associated with a widespread network encompassing frontal regions. No differences were found in these associations between hemispheres. Rey figure copy score was not associated with any mediotemporal regional cortical thickness (data not shown).

Table 1
Population baseline characteristics—The Memento cohort

| Characteristics | All participants (N = 2157) | FDG-PET participants (n = 1310) | FDG-PET performed yes versus no (*P*-value) |
|---|---|---|---|
| Median age in years, (Q1; Q3) | 71.6 (65.6; 77.1) | 72 (65.8; 77.0) | .55 |
| Female gender, n (%) | 1335 (61.9) | 770 (58.8) | .0002 |
| Educational level > 12 years, n (%) | 1172 (54.5) | 732 (56.0) | .070 |
| Number of *APOE* ε4 allele, n (%) | | | .99 |
| 0 | 1445 (70.4) | 890 (70.4) | |
| 1 | 537 (26.2) | 332 (26.2) | |
| 2 | 70 (3.4) | 43 (3.4) | |
| CDR score, n (%) | | | .018 |
| 0 | 869 (40.3) | 554 (42.3) | |
| 0.5 | 1288 (59.7) | 756 (57.7) | |
| Cognitive status, n (%) | | | .045 |
| SCD | 343 (15.9) | 219 (16.7) | |
| Pure aMCI | 196 (9.1) | 112 (8.5) | |
| Multi-domain aMCI | 924 (42.8) | 533 (40.7) | |
| Pure naMCI | 366 (17.0) | 237 (18.1) | |
| Multi-domain naMCI | 328 (15.2) | 209 (16.0) | |
| Median MMSE score, (Q1; Q3) | 28 (27; 29) | 28 (27; 29) | |
| FCSRT scores, median (Q1; Q3) | | | |
| IR (/16) | 16 (15; 16) | 16 (15; 16) | .97 |
| FR (/48) | 27 (21; 32) | 27 (21; 32) | .078 |
| ISC (/100) | 89 (77; 96) | 89 (77; 96) | .25 |

Abbreviations: CDR, clinical dementia rating; MMSE, Mini Mental State Evaluation; FCSRT, Free and Cued Selective Reminding Test; IR, immediate recall; FR, free recall; ISC, index of sensitivity to cueing; SCD, subjective cognitive decline; aMCI, amnestic mild cognitive impairment; naMCI, nonamnestic mild cognitive impairment; FDG-PET, fluorodeoxyglucose-positron emission tomography.

NOTE. Between-group comparisons were performed through $\chi^2$ test for discrete variables or analysis of variance tests for continuous variables.

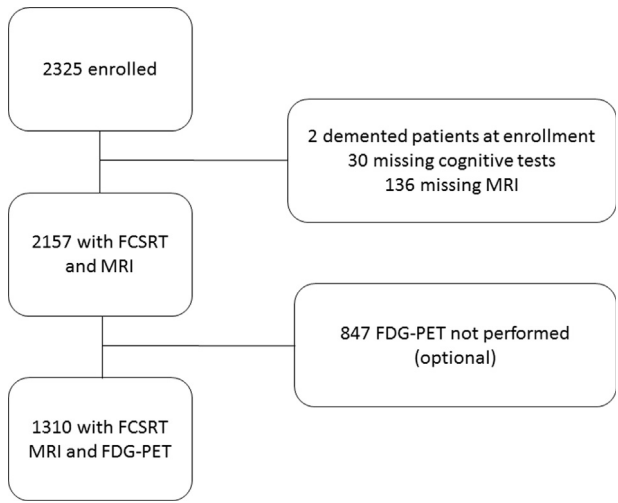

Fig. 1. Participants flow chart. Abbreviations: FCSRT, free and cued selective reminding test; FDG-PET, fluorodeoxyglucose-positron emission tomography.

In the latent class analyses (results provided in Supplementary Appendix C), IR was the subscore most strongly associated with cortical thicknesses in the superior temporal, precentral, lingual, precuneus, and latero-occipital regions. By contrast, the ISC was not associated with these structural variations, except for cortical thickness in the superior temporal cortex (coefficient = 0.10 [95% CI 0.02; 0.17]) as compared to IR (coefficient = 0.21 [95% CI 0.10; 0.31]) and FR (coefficient = 0.19 [95% CI 0.13; 0.26]).

*APOE* ε4 carriers had significantly different associations between FCSRT subscores and regional cortical thicknesses as described in Supplementary Appendices D and E. Most strikingly, the difference in median of the FR associated to the entorhinal cortex thickness was twice as important in *APOE* ε4 carriers than in noncarriers. However, none of these differences was significant after FDR correction.

### 3.2. FDG-PET correlates of FCSRT subscores

As there was no difference between the left and right hemisphere associations to FCSRT subscores, symmetrical regions were joined as metaregions of interest to study the associations to each FCSRT subscores in 47 regions (i.e., 94/2 from the AAL2 atlas, excluding the cerebellum).

The regions where the brain metabolism was significantly linked to IR and ISC scores were limited to the posterior cingulate gyri, parietotemporal junction, and medial temporal lobes albeit in a more widespread fashion for ISC than for IR (Fig. 3). Conversely, FR score was significantly related to metabolic measures in a diffuse network comprising prefrontal (medial, dorsolateral, and orbitofrontal), as well as parietal (lateral and medial) and temporal (lateral and medial) regions.

FR and ISC were significantly associated with precuneus, posterior cingulate cortex, associative parietal cortex, and temporal cortex (both left and right sides), whereas for IR, correlations were significant only in the posterior cingulate and temporal cortices, bilaterally.

The latent class analysis (Supplementary Appendix F) did not indicate singular patterns of regional metabolic association to the three FCSRT subscores. However, we found a global effect on episodic memory as a whole of the metabolic measures in 34/47 of the studied AAL regions (with the exception of the putamen, pallidum, and primary motor and sensitive areas).

Compared to noncarriers, *APOE* ε4 carriers had higher associations between FCSRT subscores and metabolism in multiple regions encompassing a large occipito-parietotemporal network for IR and FR and the same network with additional frontal and limbic regions for ISC. In contrast to the same analysis for cortical thicknesses, most of these differences remained significant after the FDR correction and are described in Supplementary Appendices G and H.

### 4. Discussion

We explored the structural (MRI) and metabolic (FDG-PET) correlates of the three main processes of episodic memory, using a cued memory test, the FCSRT, in a large cohort of elderly participants with cognitive profile ranging from SCD to MCI. We found that the three different processes involved in episodic memory are associated with different brain networks.

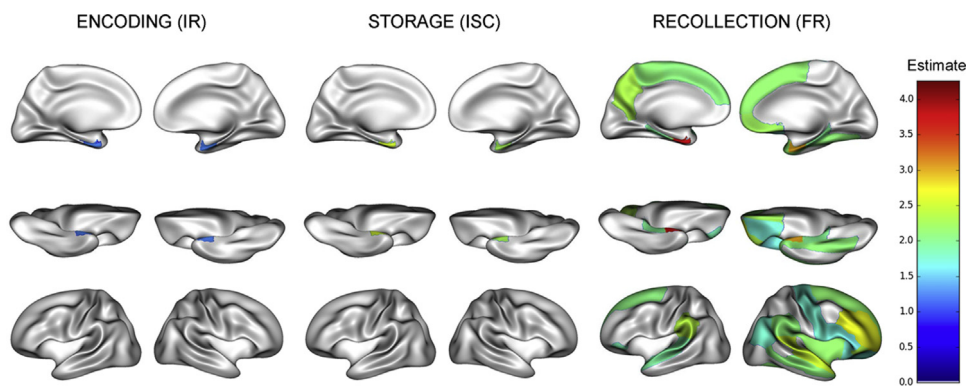

Fig. 2. Regional pattern of association between FCSRT subscores and cortical thickness (N = 2157). Abbreviations: FCSRT, free and cued selective reminding test; IR, immediate recall; ISC, index of sensitivity to cueing; FR, free recall.

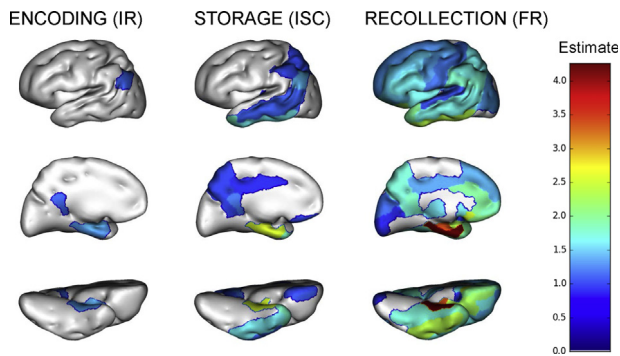

Fig. 3. Regional pattern of association between FCSRT subscores and mean regional FDG uptake values (n = 1310). Abbreviations: FCSRT, free and cued selective reminding test; FDG, fluorodeoxyglucose; ISC, index of sensitivity to cueing; FR, free recall; IR, immediate recall.

### 4.1. Validity of the FCSRT to study episodic memory

Our findings are in line with the "Attention to memory" model proposed by Cabeza and collaborators [26]. This model stipulates that the parietal cortex is involved in voluntary (top-down) retrieval of information, which is the case in the FCSRT. First, IR (which reflects the registration process) was mostly associated with metabolism in posterior brain areas. These posterior regions are associated with attentional and working memory performances [26]. Second, FR is associated with cortical thickness and metabolism in most of the brain regions and most notably in anterior brain regions. Actually, FR measures the ability to actively recollect information and is linked to executive functions. Finally, the ISC, a subscore representative of storage, is mostly associated with the cortical thickness in temporal regions [28]. This memory process is largely independent of attention and executive functions, which are impaired, for example, in pure brain vascular disease, such as in Cerebral Autosomal Dominant Arteriopathy with Subcortical Infarcts and Leukoencephalopathy [53]. In Cerebral Autosomal Dominant Arteriopathy with Subcortical Infarcts and Leukoencephalopathy, storage impairment is both rare and occurs at a later course of the disease [54].

Interestingly, the absence of differential associations according to hemispheric side for both hippocampal volumes and medial temporal lobe cortical thicknesses on FCSRT subscores is in line with previous studies [28]. In young healthy subjects, episodic memory rather involves the left hemisphere [55], but in the elderly, it involves both hemispheres. This might reflect a compensatory mechanism necessary to memorize stimuli in aging.

Our results concerning the FCSRT association with brain structures and metabolism are specific. Indeed, we did not find any similar association between the Rey's figure copy score, chosen as a nonmemory/nonlanguage cognitive process control, and the imaging markers in temporomedial regions (data not shown).

Current structural and metabolic correlates of memory exhibited some similarities. However, some regions were rather associated to a memory process on MRI (such as temporopolar regions for encoding) or FDG-PET (such as posterior cingulate cortex for storage). This is probably due both to differences in imaging acquisition and processing and to physiopathological discrepancies in MRI versus FDG-PET. The hippocampal paradox in AD (i.e., compensated metabolism that remains normal in atrophied hippocampus) is an example of such discrepancies [56]. The processing of the images relied on the use of validated pipelines and atlases that differed between the two imaging modalities, namely FreeSurfer [46] for the MRI cortical thickness ROIs and AAL2 [49] for FDG-PET ROIs. However, the macroscopic differences evidenced in our study cannot be attributed solely to the use of these different atlases. As FDG-PET was optional and performed only in a subsample, one could argue that this is the cause of the evidenced discrepancies. However, the Appendix analysis on the subsample having both MRI and FDG-PET showed the same results as in the whole group excluding a selection bias (Supplementary Appendices I and J).

A limitation in the delineation of the neural correlates of episodic memory in our study is that the FCSRT is an associative memory test with semantic cueing. Hence, some of the associated structural or metabolic regions are likely to support semantic rather than episodic processes [57]. This is for instance the case for the temporopolar association to FR. This is however the case with all verbal memory tests, especially those allowing encoding and retrieval facilitation through cueing.

### 4.2. Linking neural substrates of episodic memory to early AD diagnosis and pathology

A challenge for establishing an early AD diagnosis is to identify the pattern of memory disorder in relation to pathological injury. It has been shown that the FCSRT can quantify and qualify the memory deficits and can therefore distinguish "pure memory impairment" (failure of information storage and new memory formation) from retrieval disorders due to attention/executive changes in normal aging or frontal pathologies [11,21,22,58–60]. Such a test can identify the amnesic syndrome due to medial temporal damage that we call "the hippocampal type," and characteristically observed in AD. This syndrome is defined by poor FR and decreased total recall caused by an insufficient effect of cueing. The low performance on total recall, despite retrieval facilitation given by semantic cues, indicates poor storage capacity. Our study indicates that a decline in the retrieval process might be an earlier cognitive marker of AD than the more specific storage deficit.

When looking at the association between storage and retrieval and the regional cortical thickness, AD physiopathology distribution comes to mind.

In typical AD patients, the disease progression is stereotyped: amyloid β lesions are initially neocortical and will diffuse centripetally, and tau neurofibrillary tangles are first evidenced in the medial temporal regions before spreading in a centrifuge way [61].

The medial temporal regions underlying storage is reminiscent of the early Braak stages that can be demonstrated neuropathologically [62] or more recently by way of tau-tracer PET imaging [63]. This medial temporal involvement associated with the storage process explains the specificity of the ISC (and of the sum of the total recall) for AD even at an early (prodromal) stage and among multiple neurodegenerative conditions [19]. Conversely, the retrieval process neural substrates encompass both the medial temporal regions (affected early on during AD by the tauopathy as mentioned previously) and the regions in which amyloid deposition begins in AD (medial and dorsal prefrontal, precuneus, anterior and posterior cingulate, parietotemporal junction) [64,65]. This interpretation is in line with recent evidence suggesting that the FR score declines on average 2 years before the total recall score in cognitively healthy elderly individuals but having a positive amyloid PET scan [66].

Two conditions are considered at high risk for AD: the status of MCI and the status of *APOE* ε4 carrier. In our study, the metabolic correlates of the storage process are the same regions as those found to be hypometabolic both in MCI who rapidly progress to AD [67] and in asymptomatic *APOE* ε4 carriers [68]. This strongly suggests that FCSRT can be considered as a valuable surrogate marker of neurodegeneration in subjects at risk for AD. The fact that episodic and semantic memory processes are tested in the FCSRT explains why this test is so sensitive to early AD as both episodic and semantic impairment can be observed in this affection [69].

In AD, MRI and FDG-PET are considered valuable prognostic tools [5]. In our analytical sample, *APOE* ε4 had an impact on the degree of association between FCSRT subscores, metabolism and, to a lesser extent, cortical thickness. The stronger associations observed between structure or metabolism and FCSRT subscores in *APOE* ε4 carriers in our study is an argument to support the claim that this cognitive test can be considered, in this population of elderly SCD or MCI, as a neuropsychological prognostic marker of AD.

Among the three subscores, FR correlates to cortical thickness and metabolism in the largest cortical network (fronto-parieto-temporal associative cortices). Thus, FR is likely to decrease if any part of its associated neural network is injured. This explains why this subscore is the most sensitive in early AD. By contrast, the ISC, which is associated with cortical thickness and metabolism in the medial temporal areas is probably a more specific but less-sensitive marker. As amyloid PET imaging will be soon available for a sample of several hundred of Memento participants, it will be possible to test the hypothesis of an early cognitive impact of brain amyloidosis on the retrieval process of episodic memory. In summary, our results strengthen Wolk and Dickerson's claim that multiple measures of memory tests, underlined by different brain structures, are required to address the full spectrum of impairment that can affect AD patients [70]. These authors' work on the longitudinal follow-up of cognitively normal elderly Alzheimer's Disease Neuroimaging Initiative participants [71] confirms that cortical thickness is an early sign of AD and not only linked with cross-sectional memory impairment as demonstrated by our study but also with longitudinal cognitive decline.

The relation between FCSRT neural substrates and early-stage AD pathological patterns has direct implication for clinical care and trials. It can explain why some trials in amnestic MCI defined with the FCSRT will show some evidence of efficacy, such as the slowing of hippocampal atrophy and cortical thickness with Donepezil [72,73], whereas other trials in which MCI is not defined with the FCSRT do not, despite being more powered and longer [74,75]. Our study supports the use of the FCSRT as an important neuropsychological enrichment factor for AD in SCD and MCI trials. The FCSRT can also be used as a clinically meaningful endpoint as in the recently published INSIGHT-PreAD study [76] in which the total recall subscore dramatic decrease is used as a proxy to address the "preclinical" to "clinical" stages transition. This approach is aimed at increasing the specificity of early clinical AD detection (at the prodromal stage) to enrich clinical trial inclusions. Other studies, such as a large U.S. prevention trial Anti-Amyloid Treatment in Asymptomatic Alzheimer's Disease [77,78] and the French Multidomain Alzheimer Preventive Trial [79] have used the FCSRT as a clinical endpoint not by itself but among other tests in cognitive composite scores, namely the "Preclinical Alzheimer Cognitive Composite" score and the "MAPT-PACC" score, respectively. Although the use of composite scores allow to reduce type one error in statistical analyses, their clinical value and neural underpinnings are not as clear as individual cognitive tests (although most individual tests, and the FCSRT among them, are not purely related to one cognitive domain. In the case of the FCSRT, as mentioned previously, episodic and semantic memory processes are implicated).

### 4.3. Validity of the methodology

Strengths of our findings are related to the size of the cohort, its multimodality, and the quality of data collected for the Memento cohort, including a high degree of standardization of acquired data in all domains, from neuropsychological tests to imaging. This was organized before, during, and after (postprocessing) acquisition of data, which allowed optimal intercenter reproducibility as already described for PET imaging [48].

We acknowledge that the approach we used cannot bring the same refined information as functional MRI (fMRI) studies, which can for instance indicate which part of the hippocampus is involved in different memorization processes [36]. Our approach is complementary to fMRI delineation of the structural underpinning of memory processes and is likely to yield more robust, if less precise, results. The small number of participants included in most fMRI studies can be seen as a factor of discrepancies observed across studies (i.e., no hippocampal involvement in the retrieval of personal episodic autobiographical memory events [80] versus left hippocampal involvement [29]). The fMRI methodologies (particularly concerning the statistical analysis of results) also vary from one fMRI study to the next, and the inferences derived must be taken with caution [81]. In our study, the added value of a homogeneous population, standardized acquisition process over a relatively short interval of time, standardized quality checking, and postprocessing of data by a unique team (at the CATI [82]) and ultimately, statistical analysis taking into account both the multiple comparisons and adjustment factors allow us to draw valid conclusions from our findings. Also, the choice to study the associations of cognitive tests with predefined cortical areas derived from published atlas greatly decreases the number of statistical tests performed relatively to voxel-based comparisons while the analyzed regions remain pertinent on an anatomical and functional point of view. In any case, both types of studies are bound to provide complementary results, fMRI providing a finer delineation of subtle episodic memory functioning while our innovative methodology gives a more general and robust understanding of the major regions structurally and functionally underlying the cognitive processes of memory in aging.

This type of study has to be considered in the broader spectrum of standardized MRI postprocessing for routine clinical care. Numerous software programs are becoming available to the radiologists to help clinicians in their assessment of brain (particularly neurodegenerative) diseases [83]. This approach yet remains to be studied, but the Memento cohort seems to have the optimal design to validate it further as the participants will be followed longitudinally, allowing to determine the best marker of combination of markers to identify incipient AD or other brain diseases.

### Acknowledgments

The Memento cohort was supported by a grant from the Fondation Plan Alzheimer (Alzheimer Plan 2008-15 2012) and sponsored by the Bordeaux University Hospital. This work was also conducted by the following: CIC 1401-EC, Bordeaux University Hospital, Inserm, and Bordeaux University.

The authors acknowledge the fruitful discussion of these results with Pr Laurent Cohen, Pr Lionel Naccache, and Dr David Bendetowicz.

### Supplementary data

Supplementary data related to this article can be found at https://doi.org/10.1016/j.trci.2018.03.010.

---

### RESEARCH IN CONTEXT

1. Systematic review: We searched the literature (PubMed) for the following terms: episodic memory AND magnetic resonance imaging OR Positron emission tomography (PET) OR structural correlates OR functional correlates revealing that there was no single study nor any meta-analysis with such a large number of participants used to analyze the structural and functional correlates of episodic memory with such a high degree of clinical and imaging standardization.

2. Interpretation: Our study revealed that the free and selective reminding test and a simple and rapid association memory tests can be used to finely analyze the anatomical and functional underpinnings of episodic memory. The stronger association between regional metabolism on fluorodeoxyglucose-PET and memory performances in *APOE* ε4 carriers strengthens the diagnostic value of FCSRT for Alzheimer's disease.

3. Future directions: As amyloid PET imaging will be soon available for a sample of several hundred of Memento participants, it will be possible to test the hypothesis of an early cognitive impact of brain amyloidosis on the retrieval process of episodic memory.

---

**Appendix A. Repartition of MRI among centers and participants**

|  | Number of sites | Participants | |
|---|---|---|---|
|  |  | n | % |
| 3T MRI (n=1854) |  |  |  |
| GE MR Discovery 3T | 4 | 162 | 7.5 |
| GE Signa 3T | 6 | 127 | 5.9 |
| Philips Achieva 3T | 8 | 475 | 22.0 |
| Philips Ingenia 3T | 2 | 102 | 4.7 |
| Siemens Skyra 3T | 3 | 199 | 9.2 |
| Siemens Trio 3T | 2 | 78 | 3.6 |
| Siemens Verio 3T | 9 | 711 | 33.0 |
| 1.5T MRI (n=303) |  |  |  |
| GE Signa 1.5T | 1 | 79 | 3.7 |
| Philips Achieva 1.5T | 1 | 28 | 1.3 |
| Philips Intera 1.5T | 1 | 51 | 2.4 |
| Siemens Avanto 1.5T | 2 | 65 | 3.0 |
| Siemens Symphonytim Avanto 1.5T | 1 | 80 | 3.7 |

**Appendix B. FCSRT sub-scores distribution**

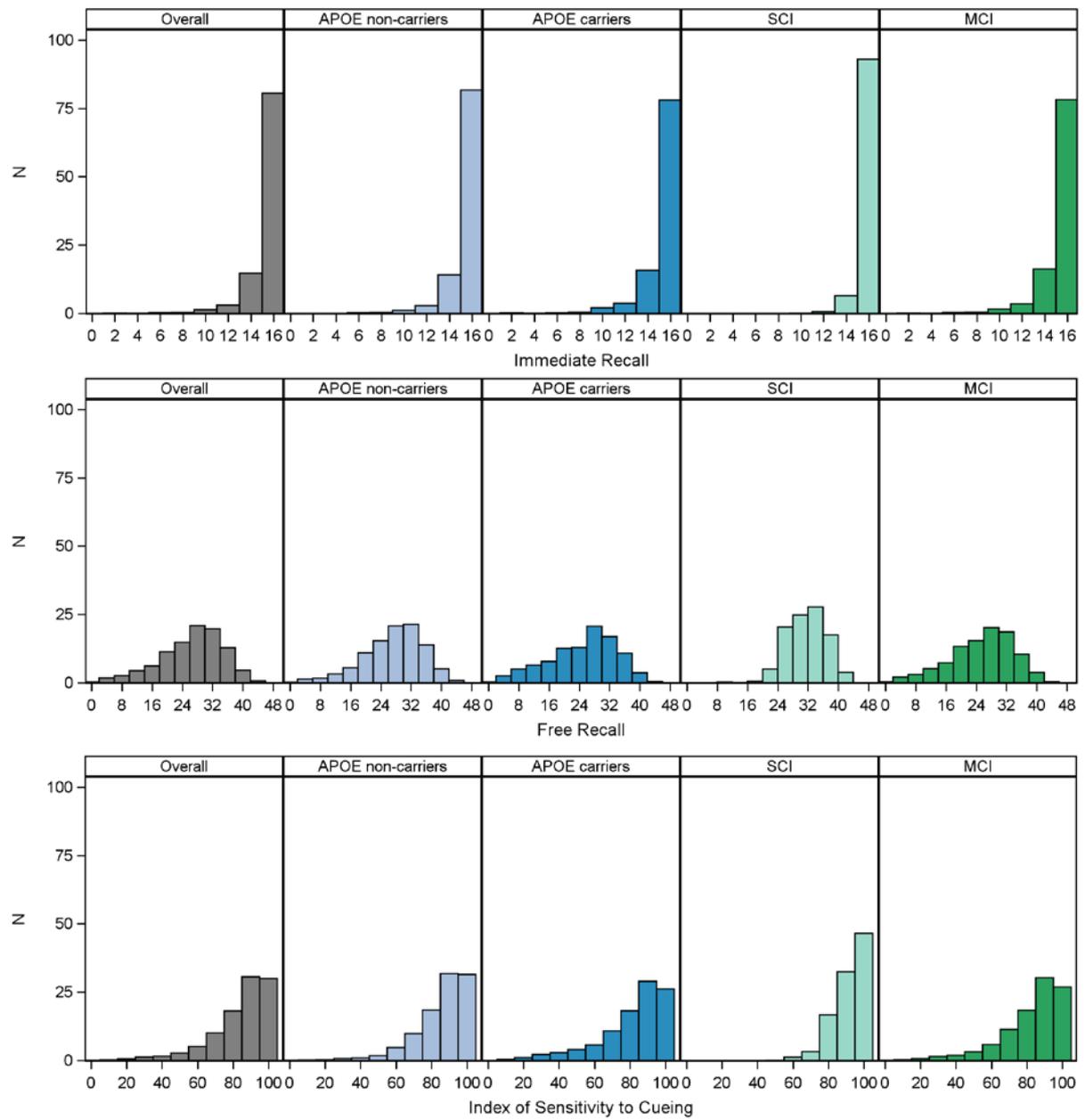

Appendix C: Latent class analysis of the associations between MRI measurements and episodic memory in the Memento Cohort

| | | | | FCSRT subscores | | |
|---|---|---|---|---|---|---|
| MRI features | Overall effect on episodic memory | FDR-corrected p values | Test of effect homogeneity (FDR-corrected p-value)) | IR | FR | ISC |
| entorhinal | 0.31 [0.24;0.37] | <0.0001 | 0.61 | | | |
| parahippocampal | 0.18 [0.12;0.24] | <0.0001 | 0.49 | | | |
| fusiform | 0.16 [0.10;0.23] | <0.0001 | 0.050 | | | |
| temporalpole | 0.14 [0.08;0.20] | <0.0001 | 0.22 | | | |
| supramarginal | 0.15 [0.09;0.21] | <0.0001 | 0.050 | | | |
| parsorbitalis | 0.07 [0.01;0.13] | 0.041 | 0.141 | | | |
| insula | 0.13 [0.07;0.19] | <0.0001 | 0.43 | | | |
| superiortemporal | 0.17 [0.10;0.23] | <0.0001 | 0.036 | 0.21 [0.10;0.31] | 0.19 [0.13;0.26] | 0.10 [0.02;0.17] |
| medialorbitofrontal | 0.10 [0.04;0.16] | 0.0017 | 0.69 | | | |
| inferiorparietal | 0.13 [0.07;0.19] | <0.0001 | 0.050 | | | |
| middletemporal | 0.13 [0.07;0.19] | 0.0001 | 0.050 | | | |
| precentral | 0.11 [0.04;0.17] | 0.0028 | 0.027 | 0.16 [0.06;0.26] | 0.13 [0.07;0.19] | 0.03 [-.04;0.10] |
| superiorfrontal | 0.07 [0.01;0.13] | 0.046 | 0.50 | | | |
| caudalmiddlefrontal | 0.09 [0.03;0.15] | 0.0057 | 0.47 | | | |
| lateralorbitofrontal | 0.08 [0.02;0.14] | 0.020 | 0.43 | | | |
| isthmuscingulate | 0.10 [0.04;0.15] | 0.0031 | 0.050 | | | |
| lingual | 0.06 [-.00;0.12] | 0.077 | 0.0048 | 0.12 [0.03;0.22] | 0.08 [0.02;0.14] | -.03 [-.10;0.04] |
| posteriorcingulate | 0.03 [-.03;0.09] | 0.30 | 0.126 | | | |
| transversetemporal | 0.08 [0.02;0.14] | 0.023 | 0.076 | | | |
| frontalpole | 0.06 [0.00;0.12] | 0.050 | 0.88 | | | |
| precuneus | 0.12 [0.06;0.18] | 0.0004 | 0.029 | 0.19 [0.09;0.29] | 0.13 [0.07;0.19] | 0.05 [-.03;0.12] |
| parsopercularis | 0.04 [-.01;0.10] | 0.18 | 0.064 | | | |
| inferiortemporal | 0.11 [0.05;0.17] | 0.0013 | 0.149 | | | |
| lateraloccipital | 0.09 [0.03;0.15] | 0.0080 | 0.029 | 0.14 [0.05;0.23] | 0.11 [0.05;0.17] | 0.02 [-.05;0.08] |
| rostralmiddlefrontal | 0.07 [0.01;0.13] | 0.042 | 0.94 | | | |

| MRI features | Overall effect on episodic memory | FDR-corrected p values | Test of effect homogeneity (FDR-corrected p-value)) | IR | FR | ISC |
|---|---|---|---|---|---|---|
| | | | **FCSRT subscores** | | | |
| postcentral | 0.09 [0.03;0.15] | 0.0049 | 0.050 | | | |
| paracentral | 0.04 [-.02;0.10] | 0.24 | 0.050 | | | |
| parstriangularis | 0.02 [-.04;0.07] | 0.63 | 0.131 | | | |
| caudalanteriorcingulate | 0.01 [-.05;0.06] | 0.84 | 0.84 | | | |
| cuneus | 0.04 [-.02;0.10] | 0.20 | 0.131 | | | |
| superiorparietal | 0.06 [0.00;0.12] | 0.054 | 0.050 | | | |
| rostralanteriorcingulat | 0.02 [-.03;0.08] | 0.44 | 0.69 | | | |
| pericalcarine | 0.02 [-.04;0.08] | 0.48 | 0.050 | | | |
| bankssts | 0.07 [0.01;0.13] | 0.042 | 0.066 | | | |

\* p-values were FDR-corrected for cortical thickness.

FCSRT: Free and Cued Selective Reminding Test , IR: Immediate Recall, FR: Free Recall,

ISC: Index of Sensitivity to Cueing , WMH: White Matter Hyperintensities.

**Appendix D. APOE interaction with cortical thickness**

Method: To determine whether APOE eps4 status could modify to relation between cortical thickness and FCSRT performance, we introduce APOE as an interaction term, and tested it to 0. For p-values < 0.05, we presented both estimations in Non APOE and APOE carriers. FDR-corrected p-values were also presented.

Part A.  Immediate Recall

|  | Global Effect | APOE interaction | APOE interaction Corrected | No APOEeps4 | APOE eps4 + |
| --- | --- | --- | --- | --- | --- |
| middletemporal | 1.14 [ 1.02; 1.26] | 0.025 | 0.63 | 1.23 [ 1.08; 1.39] | 0.97 [ 0.81; 1.16] |
| superiorparietal | 1.11 [ 1.00; 1.23] | 0.017 | 0.57 | 1.21 [ 1.07; 1.37] | 0.94 [ 0.79; 1.12] |
| precuneus | 1.14 [ 1.03; 1.26] | 0.041 | 0.70 | 1.23 [ 1.08; 1.39] | 1.00 [ 0.84; 1.18] |
| lingual | 1.07 [ 0.97; 1.18] | 0.037 | 0.70 | 1.13 [ 1.00; 1.28] | 0.91 [ 0.76; 1.08] |
| pericalcarine | 1.05 [ 0.95; 1.16] | 0.049 | 0.72 | 1.13 [ 1.00; 1.27] | 0.92 [ 0.77; 1.09] |
| cuneus | 1.01 [ 0.92; 1.12] | 0.011 | 0.56 | 1.10 [ 0.98; 1.24] | 0.85 [ 0.71; 1.01] |

Part B. Free Recall

|  | Global Effect | APOE interaction | APOE interaction Corrected | No APOEeps4 | APOE eps4 + |
| --- | --- | --- | --- | --- | --- |
| entorhinal | 4.27 [ 3.46; 5.08] | 0.0043 | 0.44 | 3.41 [ 2.40; 4.43] | 6.41 [ 4.49; 8.34] |

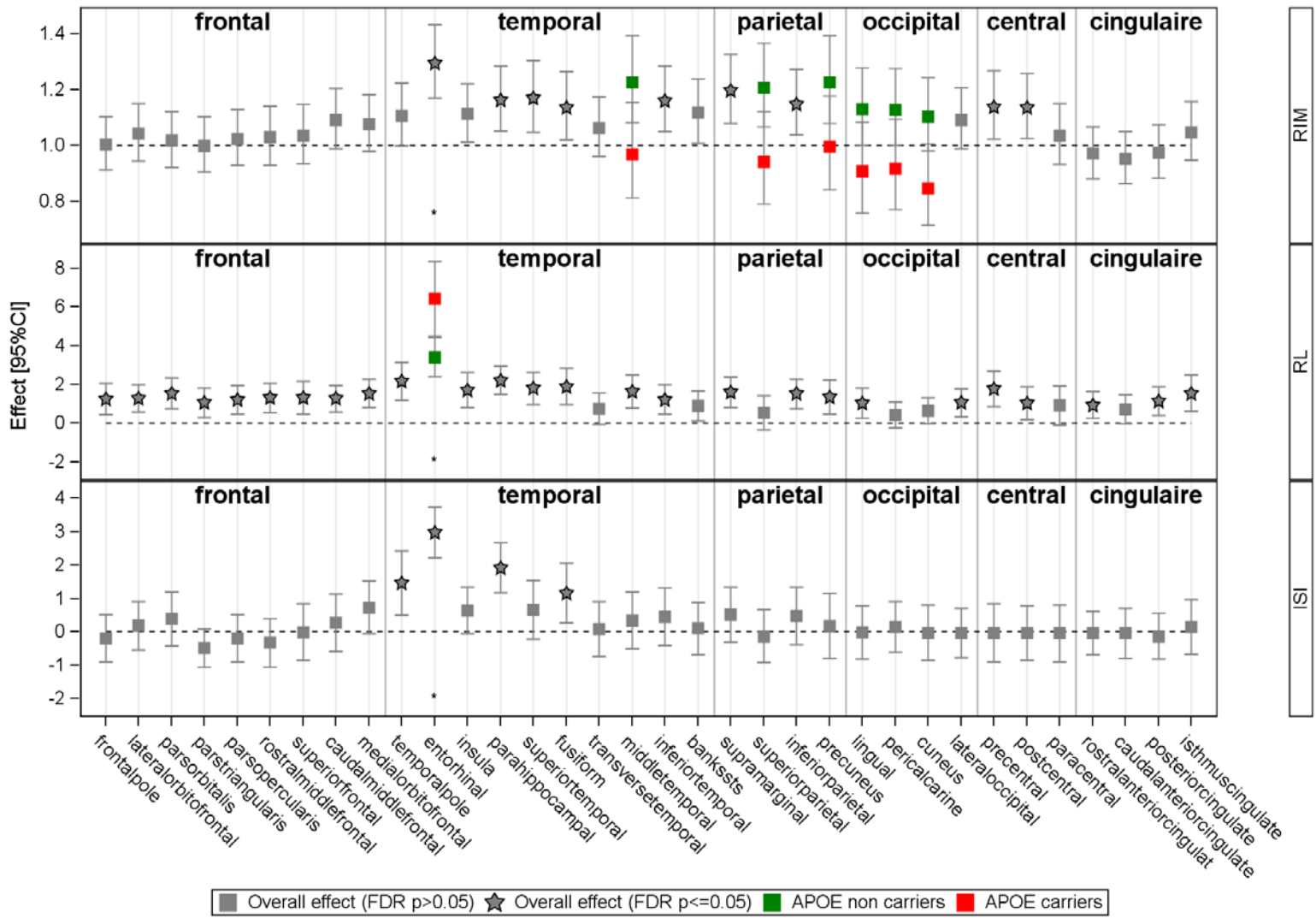

LCA APOE interaction: *** p<=0.001, ** p<=0.01, * p<=0.05

**Appendix E. Summary of interaction between cortical thickness and APOE in association with FCSRT subscores (RIM = Immediate recall, RL = Free recall, ISI = Index of sensitivity to cueing)**

Appendix F: Latent class analysis of the associations between FDG-PET Mean Uptake Values and episodic memory in the Memento Cohort

| Anatomical description | | Overall effect on episodic memory | FDR-corrected p values | Test of effect homogeneity (FDR-corrected p-value)) |
|---|---|---|---|---|
| CENTRAL | Precentral | 0.07 [-.00;0.15] | 0.084 | 1.00 |
| | Postcentral | 0.02 [-.06;0.11] | 0.61 | 1.00 |
| | Rolandic_Oper | 0.08 [0.01;0.16] | 0.043 | 0.73 |
| FRONTAL | Frontal_Sup_2 | 0.07 [0.00;0.14] | 0.058 | 0.89 |
| | Frontal_Mid_2 | 0.09 [0.03;0.15] | 0.0063 | 0.70 |
| | Frontal_Inf_Oper | 0.10 [0.04;0.17] | 0.0032 | 0.64 |
| | Frontal_Inf_Tri | 0.10 [0.04;0.16] | 0.0027 | 0.70 |
| | Frontal_Sup_Medial | 0.06 [-.00;0.13] | 0.069 | 0.70 |
| | Supp_Motor_Area | 0.03 [-.03;0.10] | 0.37 | 1.00 |
| | Paracentral_lobule | 0.03 [-.04;0.10] | 0.40 | 0.70 |
| | Frontal_Med_Orb | 0.12 [0.05;0.18] | 0.0017 | 0.64 |
| | Frontal_Inf_Orb_2 | 0.07 [0.01;0.14] | 0.040 | 0.64 |
| | Rectus | 0.15 [0.08;0.22] | 0.0002 | 0.93 |
| | OFCmed | 0.14 [0.07;0.22] | 0.0009 | 0.64 |
| | OFCant | 0.07 [0.01;0.13] | 0.029 | 0.64 |
| | OFCpost | 0.11 [0.03;0.18] | 0.0074 | 0.64 |

| Anatomical description | | Overall effect on episodic memory | FDR-corrected p values | Test of effect homogeneity (FDR-corrected p-value)) |
|---|---|---|---|---|
| | OFClat | 0.02 [-.03;0.07] | 0.46 | 0.64 |
| | Olfactory | 0.20 [0.10;0.29] | 0.0002 | 0.64 |
| TEMPORAL | Temporal_Sup | 0.10 [0.02;0.17] | 0.017 | 0.70 |
| | Heschl | 0.10 [0.04;0.15] | 0.0027 | 0.89 |
| | Temporal_Mid | 0.15 [0.07;0.22] | 0.0006 | 0.64 |
| | Temporal_Inf | 0.16 [0.07;0.24] | 0.0009 | 0.70 |
| PARIETAL | Parietal_Sup | 0.12 [0.05;0.18] | 0.0017 | 1.00 |
| | Parietal_Inf | 0.11 [0.05;0.17] | 0.0010 | 0.80 |
| | Angular | 0.16 [0.10;0.22] | <0.0001 | 0.64 |
| | SupraMarginal | 0.12 [0.05;0.19] | 0.0018 | 0.70 |
| | Precuneus | 0.11 [0.05;0.16] | 0.0012 | 0.93 |
| OCCIPITAL | Occipital_Sup | 0.04 [-.03;0.10] | 0.27 | 1.00 |
| | Occipital_Mid | 0.08 [0.02;0.15] | 0.020 | 1.00 |
| | Occipital_Inf | 0.04 [-.02;0.10] | 0.23 | 0.73 |
| | Cuneus | 0.04 [-.02;0.10] | 0.20 | 1.00 |
| | Calcarine | 0.04 [-.01;0.09] | 0.150 | 0.72 |
| | Lingual | 0.05 [-.02;0.11] | 0.16 | 1.00 |
| | Fusiform | 0.11 [0.02;0.20] | 0.023 | 0.93 |

| Anatomical description | | Overall effect on episodic memory | FDR-corrected p values | Test of effect homogeneity (FDR-corrected p-value)) |
|---|---|---|---|---|
| LIMBIC | Temporal_Pole_Sup | 0.12 [0.02;0.23] | 0.033 | 0.70 |
| | Temporal_Pole_Mid | 0.11 [0.00;0.22] | 0.067 | 0.70 |
| | Cingulate_Ant | 0.14 [0.06;0.22] | 0.0012 | 0.64 |
| | Cingulate_Mid | 0.12 [0.05;0.18] | 0.0012 | 0.93 |
| | Cingulate_Post | 0.15 [0.10;0.20] | <0.0001 | 0.64 |
| | Hippocampus | 0.41 [0.25;0.57] | <0.0001 | 0.65 |
| | ParaHippocampal | 0.35 [0.22;0.47] | <0.0001 | 0.70 |
| | Insula | 0.14 [0.05;0.22] | 0.0027 | 0.70 |
| Sub cortical grey nuclei | Amygdala | 0.26 [0.12;0.41] | 0.0014 | 0.64 |
| | Caudate | 0.12 [0.06;0.19] | 0.0009 | 0.64 |
| | Putamen | 0.05 [-.01;0.10] | 0.123 | 0.71 |
| | Pallidum | 0.04 [0.00;0.08] | 0.043 | 1.00 |
| | Thalamus | 0.10 [0.04;0.16] | 0.0032 | 0.64 |

**Appendix G. APOE interaction with 18-FDG TEP**

Part 1. Immediate Recall

| | | Global Effect | APOE interaction | APOE interaction Corrected | No APOE eps4 | APOE eps4 + |
|---|---|---|---|---|---|---|
| CENTRAL | Precentral | 0.97 [ 0.87; 1.07] | 0.50 | 0.61 | | |
| | Postcentral | 0.98 [ 0.88; 1.09] | 0.19 | 0.31 | | |
| | Rolandic_Oper | 1.07 [ 0.96; 1.18] | 0.24 | 0.37 | | |
| FRONTAL | Frontal_Sup_2 | 1.00 [ 0.92; 1.10] | 0.38 | 0.50 | | |
| | Frontal_Mid_2 | 1.02 [ 0.94; 1.11] | 0.31 | 0.44 | | |
| | Frontal_Inf_Oper | 1.07 [ 0.98; 1.17] | 0.24 | 0.37 | | |
| | Frontal_Inf_Tri | 1.06 [ 0.97; 1.15] | 0.139 | 0.25 | | |
| | Frontal_Sup_Medial | 1.01 [ 0.93; 1.11] | 0.38 | 0.51 | | |
| | Supp_Motor_Area | 0.95 [ 0.87; 1.04] | 0.35 | 0.48 | | |
| | Paracentral_lobule | 0.97 [ 0.88; 1.06] | 0.78 | 0.83 | | |
| | Frontal_Med_Orb | 1.10 [ 1.01; 1.21] | 0.110 | 0.23 | | |
| | Frontal_Inf_Orb_2 | 1.02 [ 0.93; 1.11] | 0.070 | 0.15 | | |
| | Rectus | 1.15 [ 1.04; 1.26] | 0.070 | 0.15 | | |
| | OFCmed | 1.08 [ 0.97; 1.19] | 0.0091 | 0.042 | 0.98 [ 0.87; 1.10] | 1.30 [ 1.08; 1.57] |
| | OFCant | 1.02 [ 0.94; 1.10] | 0.027 | 0.088 | 0.96 [ 0.88; 1.06] | 1.17 [ 1.00; 1.36] |
| | OFCpost | 1.06 [ 0.96; 1.17] | 0.032 | 0.094 | 0.99 [ 0.88; 1.11] | 1.23 [ 1.03; 1.47] |
| | OFClat | 0.95 [ 0.88; 1.02] | 0.044 | 0.118 | 0.90 [ 0.83; 0.99] | 1.05 [ 0.92; 1.21] |
| | Olfactory | 1.16 [ 1.02; 1.32] | 0.45 | 0.57 | | |
| TEMPORAL | Temporal_Sup | 1.11 [ 1.00; 1.23] | 0.063 | 0.150 | | |
| | Heschl | 1.07 [ 0.99; 1.16] | 0.20 | 0.32 | | |
| | Temporal_Mid | 1.13 [ 1.02; 1.25] | 0.0028 | 0.024 | 1.01 [ 0.89; 1.14] | 1.38 [ 1.15; 1.64] |
| | Temporal_Inf | 1.13 [ 1.02; 1.27] | 0.0037 | 0.027 | 1.01 [ 0.89; 1.16] | 1.41 [ 1.16; 1.71] |
| PARIETAL | Parietal_Sup | 1.03 [ 0.94; 1.14] | 0.014 | 0.057 | 0.96 [ 0.86; 1.07] | 1.25 [ 1.04; 1.50] |
| | Parietal_Inf | 1.07 [ 0.99; 1.16] | 0.021 | 0.073 | 1.00 [ 0.91; 1.10] | 1.22 [ 1.06; 1.40] |
| | Angular | 1.12 [ 1.04; 1.21] | 0.0029 | 0.024 | 1.04 [ 0.95; 1.13] | 1.31 [ 1.14; 1.50] |
| | SupraMarginal | 1.14 [ 1.03; 1.25] | 0.019 | 0.070 | 1.05 [ 0.94; 1.17] | 1.33 [ 1.12; 1.59] |
| | Precuneus | 1.06 [ 0.98; 1.15] | 0.013 | 0.055 | 0.99 [ 0.90; 1.09] | 1.22 [ 1.06; 1.41] |
| OCCIPITAL | Occipital_Sup | 1.01 [ 0.93; 1.11] | 0.022 | 0.077 | 0.95 [ 0.86; 1.06] | 1.18 [ 1.00; 1.38] |
| | Occipital_Mid | 1.07 [ 0.98; 1.17] | 0.019 | 0.070 | 0.99 [ 0.89; 1.10] | 1.23 [ 1.05; 1.44] |
| | Occipital_Inf | 1.01 [ 0.93; 1.10] | 0.0054 | 0.030 | 0.92 [ 0.83; 1.02] | 1.19 [ 1.02; 1.38] |
| | Cuneus | 1.04 [ 0.95; 1.12] | 0.0071 | 0.036 | 0.97 [ 0.88; 1.06] | 1.22 [ 1.05; 1.42] |
| | Calcarine | 1.01 [ 0.94; 1.09] | 0.057 | 0.138 | | |

|  |  | Global Effect | APOE interaction | APOE interaction Corrected | No APOE eps4 | APOE eps4 + |
|---|---|---|---|---|---|---|
|  | Lingual | 1.01 [ 0.92; 1.11] | 0.103 | 0.21 |  |  |
|  | Fusiform | 1.11 [ 0.98; 1.25] | 0.018 | 0.069 | 1.00 [ 0.86; 1.16] | 1.36 [ 1.09; 1.70] |
| LIMBIC | Temporal_Pole_Sup | 1.20 [ 1.03; 1.39] | 0.29 | 0.43 |  |  |
|  | Temporal_Pole_Mid | 1.12 [ 0.96; 1.30] | 0.018 | 0.069 | 0.98 [ 0.82; 1.18] | 1.41 [ 1.09; 1.83] |
|  | Cingulate_Ant | 1.11 [ 1.00; 1.23] | 0.119 | 0.23 |  |  |
|  | Cingulate_Mid | 1.09 [ 1.00; 1.19] | 0.083 | 0.18 |  |  |
|  | Cingulate_Post | 1.14 [ 1.07; 1.21] | 0.030 | 0.090 | 1.09 [ 1.01; 1.17] | 1.25 [ 1.12; 1.40] |
|  | Hippocampus | 1.46 [ 1.18; 1.80] | 0.24 | 0.37 |  |  |
|  | ParaHippocampal | 1.36 [ 1.15; 1.61] | 0.070 | 0.15 |  |  |
|  | Insula | 1.10 [ 0.98; 1.23] | 0.70 | 0.79 |  |  |
| SCGN | Amygdala | 1.44 [ 1.19; 1.74] | 0.47 | 0.59 |  |  |
|  | Caudate | 1.05 [ 0.97; 1.15] | 0.59 | 0.71 |  |  |
|  | Putamen | 0.99 [ 0.91; 1.06] | 0.85 | 0.89 |  |  |
|  | Pallidum | 1.00 [ 0.95; 1.05] | 0.121 | 0.23 |  |  |
|  | Thalamus | 1.03 [ 0.94; 1.12] | 0.56 | 0.67 |  |  |

## Part 2. Free Recall

|  |  | Global Effect | APOE interaction | APOE interaction Corrected | No APOE eps4 | APOE eps4 + |
|---|---|---|---|---|---|---|
| CENTRAL | Postcentral | 0.86 [ 0.06; 1.66] | 0.051 | 0.129 |  |  |
|  | Precentral | 0.98 [ 0.14; 1.81] | 0.086 | 0.18 |  |  |
|  | Rolandic_Oper | 1.11 [ 0.47; 1.75] | 0.18 | 0.31 |  |  |
| FRONTAL | Supp_Motor_Area | 0.63 [-0.12; 1.37] | 0.089 | 0.19 |  |  |
|  | Frontal_Inf_Oper | 1.19 [ 0.49; 1.89] | 0.048 | 0.124 | 1.06 [ 0.20; 1.91] | 2.74 [ 1.33; 4.15] |
|  | OFClat | 1.29 [ 0.68; 1.89] | 0.057 | 0.138 |  |  |
|  | Frontal_Sup_2 | 1.30 [ 0.61; 1.99] | 0.048 | 0.124 | 0.87 [ 0.14; 1.60] | 2.51 [ 1.00; 4.02] |
|  | Frontal_Sup_Medial | 1.40 [ 0.75; 2.04] | 0.23 | 0.37 |  |  |
|  | Frontal_Inf_Orb_2 | 1.42 [ 0.69; 2.14] | 0.42 | 0.54 |  |  |
|  | Frontal_Mid_2 | 1.55 [ 0.93; 2.17] | 0.017 | 0.065 | 1.27 [ 0.55; 1.99] | 3.16 [ 2.00; 4.31] |
|  | Frontal_Inf_Tri | 1.58 [ 0.87; 2.29] | 0.113 | 0.23 |  |  |
|  | OFCant | 1.69 [ 1.00; 2.38] | 0.30 | 0.44 |  |  |
|  | Frontal_Med_Orb | 1.79 [ 1.09; 2.49] | 0.121 | 0.23 |  |  |
|  | OFCpost | 2.11 [ 1.35; 2.87] | 0.15 | 0.27 |  |  |
|  | Rectus | 2.26 [ 1.56; 2.96] | 0.32 | 0.45 |  |  |

| | | Global Effect | APOE interaction | APOE interaction Corrected | No APOE eps4 | APOE eps4 + |
|---|---|---|---|---|---|---|
| | OFCmed | 2.30 [ 1.50; 3.11] | 0.029 | 0.089 | 1.62 [ 0.55; 2.69] | 4.17 [ 2.46; 5.88] |
| | Olfactory | 2.44 [ 1.43; 3.45] | 0.44 | 0.56 | | |
| | Paracentral_lobule | -0.15 [-0.83; 0.53] | 0.083 | 0.18 | | |
| TEMPORAL | Heschl | 0.65 [ 0.09; 1.22] | 0.42 | 0.54 | | |
| | Temporal_Sup | 0.94 [ 0.15; 1.74] | 0.028 | 0.088 | 0.21 [-0.66; 1.09] | 2.37 [ 0.62; 4.11] |
| | Temporal_Mid | 1.63 [ 0.74; 2.51] | 0.0016 | 0.020 | 0.64 [-0.32; 1.61] | 4.05 [ 2.32; 5.78] |
| | Temporal_Inf | 2.09 [ 1.20; 2.99] | 0.0002 | 0.011 | 1.13 [ 0.13; 2.14] | 4.40 [ 2.74; 6.05] |
| PARIETAL | SupraMarginal | 1.44 [ 0.71; 2.18] | 0.025 | 0.082 | 1.11 [ 0.19; 2.03] | 3.31 [ 1.75; 4.87] |
| | Parietal_Sup | 1.63 [ 0.83; 2.43] | 0.0004 | 0.011 | 0.79 [-0.03; 1.60] | 4.14 [ 2.66; 5.63] |
| | Parietal_Inf | 1.63 [ 0.98; 2.27] | 0.0057 | 0.031 | 1.15 [ 0.48; 1.81] | 3.48 [ 2.24; 4.71] |
| | Precuneus | 1.77 [ 1.15; 2.40] | 0.0034 | 0.027 | 1.05 [ 0.28; 1.83] | 3.50 [ 2.00; 5.00] |
| | Angular | 1.79 [ 1.10; 2.47] | 0.0003 | 0.011 | 1.28 [ 0.57; 2.00] | 4.03 [ 2.92; 5.14] |
| OCCIPITAL | Occipital_Inf | 0.65 [-0.02; 1.32] | 0.0004 | 0.011 | -0.09 [-0.87; 0.69] | 2.79 [ 1.33; 4.25] |
| | Occipital_Sup | 0.69 [-0.07; 1.44] | 0.0003 | 0.011 | -0.10 [-0.97; 0.78] | 3.62 [ 1.92; 5.32] |
| | Lingual | 0.72 [-0.12; 1.56] | 0.015 | 0.060 | -0.00 [-0.92; 0.92] | 2.17 [ 0.62; 3.72] |
| | Calcarine | 0.83 [ 0.20; 1.46] | 0.015 | 0.060 | 0.29 [-0.35; 0.94] | 2.10 [ 1.00; 3.20] |
| | Cuneus | 0.98 [ 0.27; 1.69] | 0.0013 | 0.020 | 0.37 [-0.38; 1.12] | 3.28 [ 1.74; 4.83] |
| | Occipital_Mid | 1.15 [ 0.33; 1.96] | 0.0024 | 0.023 | 0.50 [-0.24; 1.24] | 3.31 [ 1.76; 4.85] |
| | Fusiform | 1.40 [ 0.19; 2.61] | 0.028 | 0.088 | 0.80 [-0.31; 1.92] | 3.74 [ 1.87; 5.60] |
| LIMBIC | Cingulate_Mid | 1.47 [ 0.79; 2.15] | 0.120 | 0.23 | | |
| | Temporal_Pole_Mid | 1.52 [ 0.28; 2.76] | 0.0018 | 0.022 | 0.71 [-0.65; 2.06] | 4.71 [ 2.66; 6.76] |
| | Temporal_Pole_Sup | 1.70 [ 0.47; 2.93] | 0.056 | 0.137 | | |
| | Insula | 1.76 [ 0.81; 2.71] | 0.94 | 0.96 | | |
| | Cingulate_Ant | 1.80 [ 0.97; 2.63] | 0.96 | 0.97 | | |
| | Cingulate_Post | 2.29 [ 1.59; 2.99] | 0.022 | 0.075 | 1.44 [ 0.71; 2.16] | 3.09 [ 1.76; 4.42] |
| | Hippocampus | 3.23 [ 1.53; 4.93] | 0.72 | 0.80 | | |
| | ParaHippocampal | 3.70 [ 2.05; 5.34] | 0.0014 | 0.020 | 2.39 [ 0.93; 3.85] | 7.26 [ 4.74; 9.78] |
| SCGN | Pallidum | 0.40 [ 0.01; 0.79] | 0.64 | 0.74 | | |
| | Putamen | 0.63 [ 0.04; 1.21] | 0.93 | 0.96 | | |
| | Caudate | 1.55 [ 0.90; 2.20] | 0.38 | 0.50 | | |
| | Thalamus | 1.80 [ 1.19; 2.41] | 0.62 | 0.73 | | |
| | Amygdala | 2.32 [ 0.94; 3.69] | 0.28 | 0.41 | | |

Part 3. Index of Sensitivity to Cueing

| | | Global Effect | APOE interaction | APOE interaction Corrected | No APOE eps4 | APOE eps4 + |
|---|---|---|---|---|---|---|
| CENTRAL | Postcentral | 0.67 [-0.09; 1.43] | 0.115 | 0.23 | | |
| | Precentral | 0.77 [ 0.08; 1.46] | 0.135 | 0.25 | | |
| | Rolandic_Oper | 0.99 [ 0.15; 1.83] | 0.040 | 0.114 | 0.25 [-0.53; 1.03] | 2.22 [ 0.59; 3.85] |
| FRONTAL | Paracentral_lobule | 0.00 [-0.59; 0.60] | 0.46 | 0.58 | | |
| | Supp_Motor_Area | 0.31 [-0.22; 0.84] | 0.118 | 0.23 | | |
| | OFClat | 0.55 [-0.06; 1.15] | 0.0050 | 0.030 | 0.09 [-0.48; 0.66] | 1.93 [ 0.72; 3.15] |
| | Frontal_Sup_Medial | 0.60 [ 0.01; 1.18] | 0.043 | 0.118 | 0.21 [-0.43; 0.85] | 1.69 [ 0.36; 3.02] |
| | Frontal_Inf_Orb_2 | 0.70 [ 0.10; 1.30] | 0.0023 | 0.023 | 0.28 [-0.34; 0.90] | 2.45 [ 1.18; 3.72] |
| | OFCant | 0.74 [ 0.20; 1.29] | 0.0084 | 0.041 | 0.36 [-0.25; 0.98] | 1.99 [ 0.87; 3.10] |
| | Frontal_Sup_2 | 0.79 [ 0.25; 1.33] | 0.0072 | 0.036 | 0.42 [-0.29; 1.12] | 2.36 [ 0.97; 3.75] |
| | Frontal_Mid_2 | 0.84 [ 0.23; 1.44] | 0.0004 | 0.011 | 0.38 [-0.24; 1.00] | 2.36 [ 1.32; 3.41] |
| | OFCpost | 0.93 [ 0.18; 1.69] | 0.0021 | 0.023 | 0.28 [-0.49; 1.05] | 2.69 [ 1.24; 4.14] |
| | Frontal_Inf_Oper | 0.95 [ 0.32; 1.58] | 0.0051 | 0.030 | 0.52 [-0.16; 1.20] | 2.56 [ 1.22; 3.91] |
| | Frontal_Inf_Tri | 0.95 [ 0.34; 1.56] | 0.0008 | 0.015 | 0.30 [-0.34; 0.95] | 2.58 [ 1.38; 3.77] |
| | Olfactory | 1.04 [ 0.03; 2.05] | 0.26 | 0.39 | | |
| | Frontal_Med_Orb | 1.05 [ 0.37; 1.73] | 0.0004 | 0.011 | 0.46 [-0.33; 1.25] | 3.00 [ 1.66; 4.35] |
| | OFCmed | 1.30 [ 0.47; 2.13] | 0.0009 | 0.016 | 0.46 [-0.37; 1.30] | 3.16 [ 1.72; 4.60] |
| | Rectus | 1.36 [ 0.64; 2.09] | 0.0043 | 0.029 | 0.55 [-0.23; 1.33] | 2.92 [ 1.46; 4.38] |
| TEMPORAL | Heschl | 0.97 [ 0.35; 1.59] | 0.25 | 0.38 | | |
| | Temporal_Sup | 1.09 [ 0.31; 1.87] | 0.033 | 0.095 | 0.48 [-0.44; 1.39] | 2.28 [ 0.87; 3.69] |
| | Temporal_Mid | 1.33 [ 0.51; 2.15] | 0.0007 | 0.015 | 0.39 [-0.45; 1.23] | 3.01 [ 1.59; 4.44] |
| | Temporal_Inf | 1.58 [ 0.77; 2.39] | 0.011 | 0.047 | 0.83 [-0.22; 1.88] | 3.23 [ 1.80; 4.67] |
| PARIETAL | Precuneus | 0.99 [ 0.35; 1.63] | 0.041 | 0.114 | 0.46 [-0.24; 1.16] | 2.04 [ 0.68; 3.39] |
| | Parietal_Sup | 1.10 [ 0.34; 1.86] | 0.0065 | 0.035 | 0.48 [-0.25; 1.22] | 3.01 [ 1.32; 4.70] |
| | SupraMarginal | 1.17 [ 0.37; 1.97] | 0.0095 | 0.043 | 0.43 [-0.38; 1.24] | 2.56 [ 1.10; 4.02] |
| | Parietal_Inf | 1.22 [ 0.62; 1.82] | 0.046 | 0.124 | 0.64 [-0.03; 1.31] | 2.11 [ 0.85; 3.37] |
| | Angular | 1.47 [ 0.88; 2.06] | 0.0037 | 0.027 | 0.75 [ 0.04; 1.45] | 2.68 [ 1.51; 3.86] |
| OCCIPITAL | Calcarine | 0.55 [-0.01; 1.11] | 0.031 | 0.092 | 0.26 [-0.36; 0.88] | 1.52 [ 0.35; 2.69] |
| | Lingual | 0.72 [ 0.04; 1.39] | 0.126 | 0.23 | | |
| | Cuneus | 0.79 [ 0.19; 1.40] | 0.067 | 0.15 | | |
| | Occipital_Sup | 0.88 [ 0.19; 1.58] | 0.0067 | 0.035 | 0.27 [-0.58; 1.12] | 2.35 [ 0.93; 3.77] |
| | Occipital_Inf | 0.97 [ 0.33; 1.61] | 0.0046 | 0.030 | 0.29 [-0.49; 1.06] | 2.38 [ 1.04; 3.73] |
| | Occipital_Mid | 1.04 [ 0.38; 1.71] | 0.055 | 0.136 | | |
| | Fusiform | 1.30 [ 0.42; 2.18] | 0.0043 | 0.029 | 0.50 [-0.40; 1.39] | 3.43 [ 1.73; 5.12] |
| LIMBIC | Cingulate_Ant | 0.97 [ 0.20; 1.73] | 0.17 | 0.29 | | |

|  |  | Global Effect | APOE interaction | APOE interaction Corrected | No APOE eps4 | APOE eps4 + |
|---|---|---|---|---|---|---|
| SCGN | Insula | 1.04 [ 0.18; 1.90] | 0.44 | 0.56 |  |  |
|  | Cingulate_Post | 1.15 [ 0.62; 1.68] | 0.0029 | 0.024 | 0.68 [ 0.13; 1.24] | 2.14 [ 1.30; 2.98] |
|  | Cingulate_Mid | 1.17 [ 0.50; 1.83] | 0.023 | 0.079 | 0.53 [-0.26; 1.33] | 2.19 [ 0.90; 3.48] |
|  | Temporal_Pole_Sup | 1.65 [ 0.59; 2.72] | 0.0024 | 0.023 | 0.43 [-0.74; 1.60] | 3.78 [ 2.01; 5.54] |
|  | Temporal_Pole_Mid | 1.98 [ 0.88; 3.08] | 0.0005 | 0.013 | 0.50 [-0.83; 1.82] | 4.39 [ 2.58; 6.20] |
|  | ParaHippocampal | 2.59 [ 1.42; 3.77] | 0.0015 | 0.020 | 1.13 [-0.32; 2.57] | 5.44 [ 3.19; 7.68] |
|  | Hippocampus | 2.90 [ 1.03; 4.76] | 0.0096 | 0.043 | 1.22 [-0.78; 3.22] | 5.41 [ 2.36; 8.47] |
|  | Putamen | 0.33 [-0.18; 0.85] | 0.34 | 0.47 |  |  |
|  | Pallidum | 0.34 [-0.05; 0.72] | 0.0037 | 0.027 | 0.08 [-0.32; 0.48] | 1.43 [ 0.64; 2.22] |
|  | Thalamus | 0.58 [-0.09; 1.24] | 0.64 | 0.74 |  |  |
|  | Caudate | 0.80 [ 0.09; 1.50] | 0.31 | 0.44 |  |  |
|  | Amygdala | 1.91 [ 0.23; 3.59] | 0.0053 | 0.030 | 0.48 [-1.00; 1.96] | 4.89 [ 2.18; 7.61] |

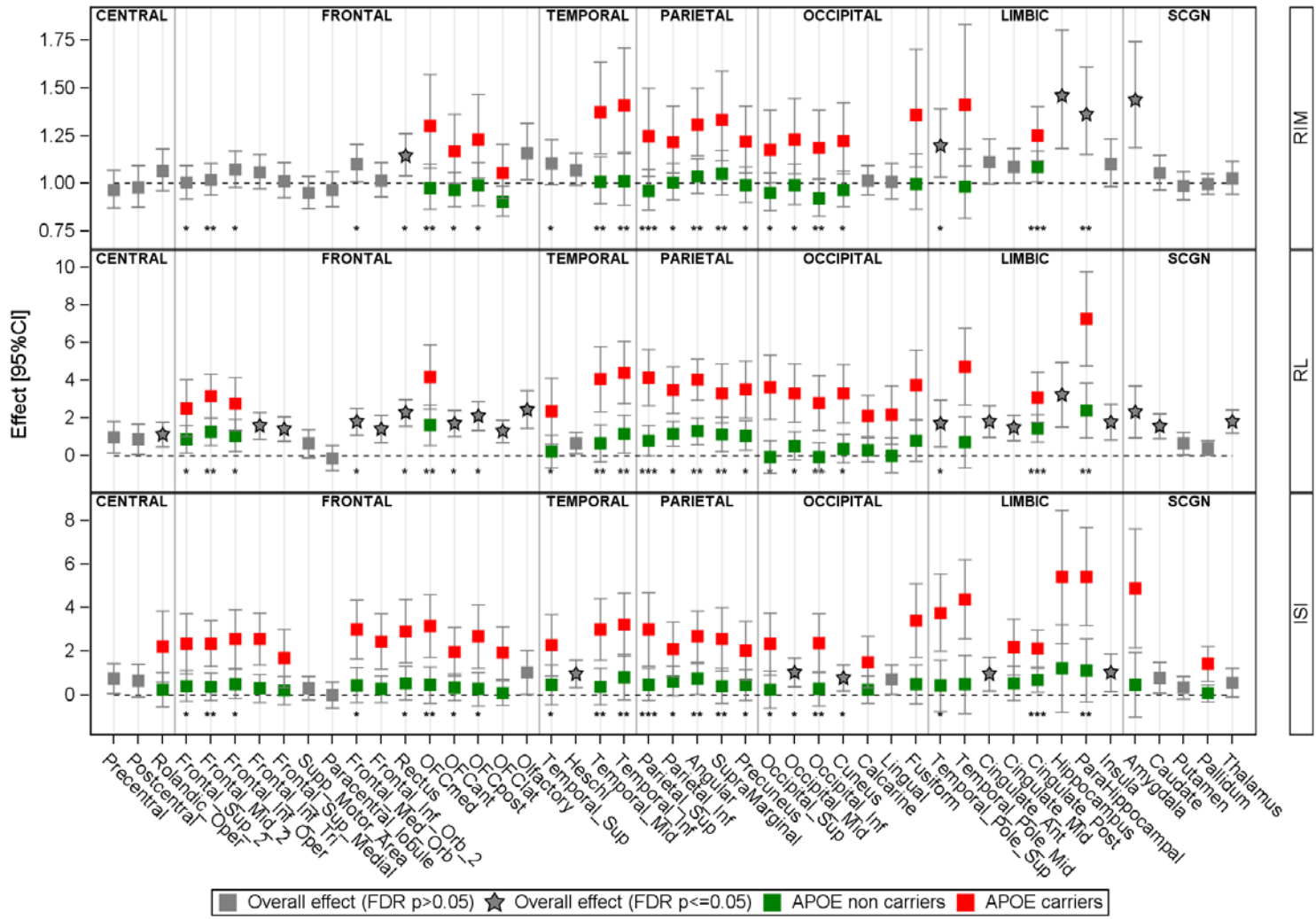

LCA APOE interaction: *** p<=0.001, ** p<=0.01, * p<=0.05

Legend: Overall effect (FDR p>0.05) ☆ Overall effect (FDR p<=0.05) ■ APOE non carriers ■ APOE carriers

**Appendix H. Summary of interaction between regional metabolism and APOE in association with FCSRT subscores (RIM = Immediate recall, RL = Free recall, ISI = Index of sensitivity to cueing)**

Appendix I: Visual comparison of the regional patterns of association between FCSRT subscores (RIM = Immediate recall, RL = Free recall, ISI = Index of sensitivity to cueing) and cortical thickness in the whole group (N=2157) and the MRI+FDG-PET subsample cohort (n=1310). Regions name are explicated in Appendix J

Appendix J: Freesurfer regions in which the cortical thickness was measured

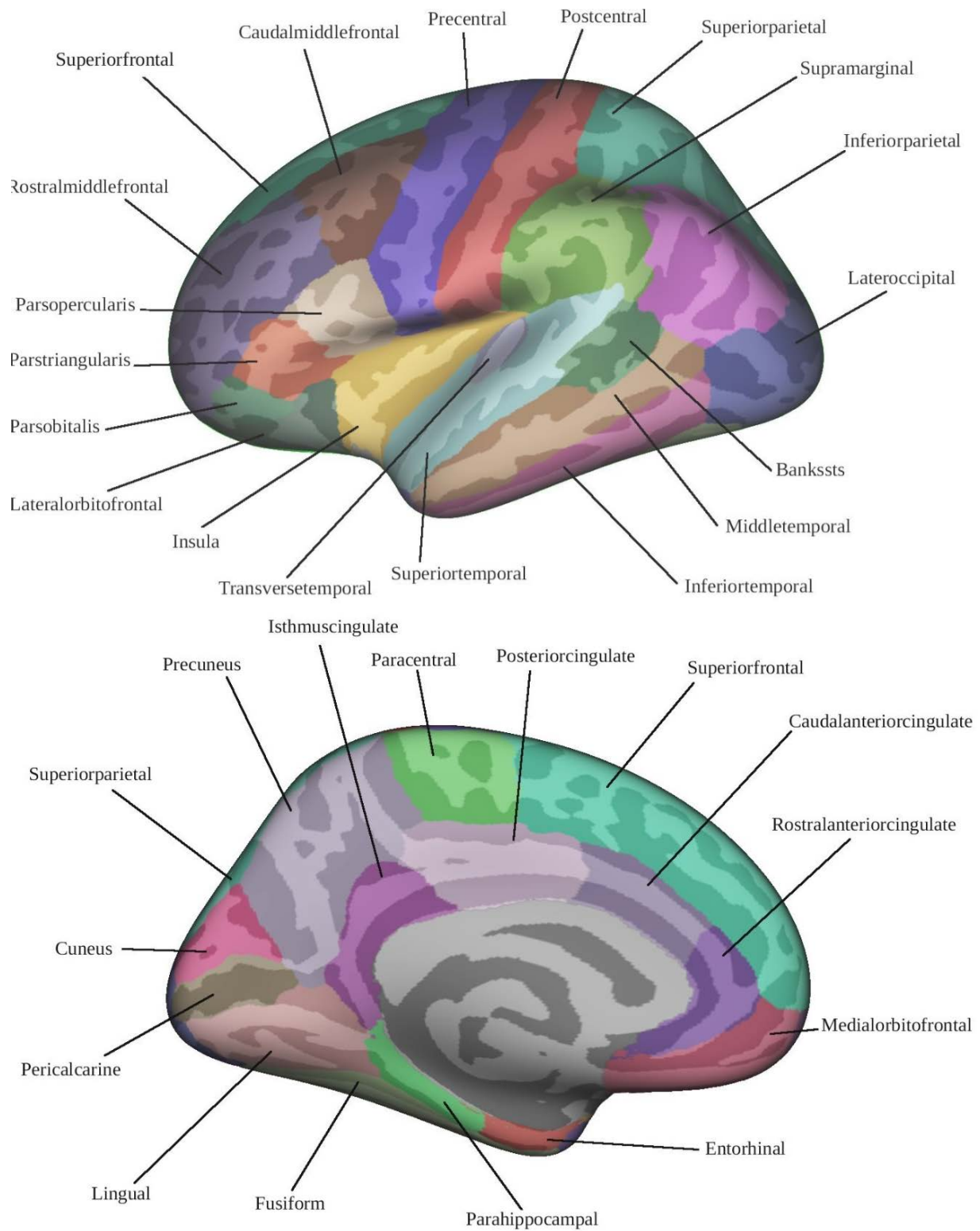

Bertrand, Accart, Memory Resource and Research Centre of Lille, CHRU de Lille, Lille, France

Geneviève, Achdjibachdian, Memory Resource and Research Centre of Marseille, CHU de Marseille, Marseille, France

Isabelle, Addra, CIC-1401 Clinical Epidemiology, CHU Bordeaux, Bordeaux, France

Sarah, Adjal, Memory Resource and Research Centre of Bordeaux, CHU Bordeaux, Bordeaux, France

Timothée, Albasser, Memory Resource and Research Centre of Strasbourg, CHRU de Strasbourg, Strasbourg, France

Michèle, Allard, Memory Resource and Research Centre of Bordeaux, CHU de Bordeaux, Bordeaux France

Sandrine, Andrieu, Memory Resource and Research Centre of Toulouse, CHU de Toulouse, Toulouse, France

Cédric, Annweiller, Memory Resource and Research Centre of Angers, CHU d'Angers, Angers, France

Pierre, Anthony, Memory Resource and Research Centre of Colmar, Colmar, France

Elise, Antoine, Memory Resource and Research Centre of Lyon, Hospices Civils de Lyon, Lyon, France

Jean-Paul, Armspach, Memory Resource and Research Centre of Strasbourg, CHRU de Strasbourg, Strasbourg, France

Christine, Astier, Memory Resource and Research Centre of Strasbourg, CHRU de Strasbourg, Strasbourg, France

Vanessa, Auberti, Memory Resource and Research Centre of Bordeaux, CHU de Bordeaux, Bordeaux, France

Christelle, Audrain, Institute of Memory and Alzheimer's Disease (IM2A), France and Brain and Spine Institute (ICM), France UMR S 1127, Department of Neurology, AP-HP, Pitié-Salpêtrière University Hospital, Sorbonne Universities, Pierre et Marie Curie University, Paris, France

Alexandre, Augier, Memory Clinic of Avicenne, Hôpital Avicenne, Bobigny, France

Sophie, Auriacombe, Memory Resource and Research Centre of Bordeaux, CHU de Bordeaux, Bordeaux, France

John, Avet, Memory Resource and Research Centre of Saint-Etienne, CHU de Saint-Etienne, Saint-Etienne, France

Romain, Bachelet, Memory Resource and Research Centre of Lyon, Hospices Civils de Lyon, Lyon, France

Olivier, Bailon, Memory Clinic of Avicenne, Hôpital Avicenne, Bobigny, France

Hélène, Bansard, Memory Resource and Research Centre of Tours, CHRU de Tours, Tours,  France

Laurent, Baranti, Memory Resource and Research Centre of Tours, CHRU de Tours, Tours,  France

Fabrice-Guy, Barral, Memory Resource and Research Centre of Saint-Etienne, CHU de Saint-Etienne, Saint-Etienne, France

Jean, Barré, Memory Resource and Research Centre of Angers, CHU d'Angers, Angers, France

Yoan, Barsznica, Memory Resource and Research Centre of Besançon, Besançon, France

Annick, Barthelaix, Memory Resource and Research Centre of Angers, CHU d'Angers, Angers, France

Laurie, Barthelemi, Memory Resource and Research Centre of Montpellier, CHU de Montpellier, Montpellier, France

Fanny, Barthelemy, Memory Resource and Research Centre of Marseille, CHU de Marseille, Marseille, France

Anthony, Bathsavanis, Memory Resource and Research Centre of Lyon, Hospices Civils de Lyon, Lyon, France

Vanessa, Baudiffier, Memory Resource and Research Centre of Poitiers, CHU de Poitiers, Poitiers, France

Sophie, Bayer, Memory Resource and Research Centre of Strasbourg, CHRU de Strasbourg, Strasbourg, France

Catherine, Bayle, Memory Resource and Research Centre of Paris Broca, AP-HP, Paris, France

Mélinda, Beaudenon, Memory Resource and Research Centre of Angers, CHU d'Angers, Angers, France

Emilie, Beaufils, Memory Resource and Research Centre of Tours, CHRU de Tours, Tours,  France

Yannick, Bejot, Memory Resource and Research Centre of Dijon, CHU Dijon Bourgogne, Dijon, France

Sandrine, Belkadi, Memory Resource and Research Centre of Poitiers, CHU de Poitiers, Poitiers, France

Julie, Bellet, Memory Resource and Research Centre of Lille, CHRU de Lille, Lille, France

Marwa, Ben, Yacoub, Memory Clinic of Avicenne, Hôpital Avicenne, Bobigny, France

Habib, Benali, Institute of Memory and Alzheimer's Disease (IM2A), France and Brain and Spine Institute (ICM), France UMR S 1127, Department of Neurology, AP-HP, Pitié-Salpêtrière University Hospital, Sorbonne Universities, Pierre et Marie Curie University, Paris, France

Karim, Bennys, Memory Resource and Research Centre of Montpellier, CHU de Montpellier, Montpellier, France

Nadine, Bensoussan, Memory Resource and Research Centre of Marseille, CHU de Marseille, Marseille, France

Géraldine, Bera, Laboratoire d'Imagerie Biomédicale, Sorbonne Universités, UPMC Univ Paris 06, Inserm U 1146, CNRS UMR 7371, F-75006 Paris, France NeuroSpin, I2BM, Commissariat à l'Energie Atomique, France

Eric, Berger, Memory Resource and Research Centre of Besançon, Besançon, France

Juliette, Berger, Memory Resource and Research Centre of Clermont-Ferrand, CHU de Clermont-Ferrand, Clermont-Ferrand, France

Marc, G, Berger, Memory Resource and Research Centre of Clermont-Ferrand, CHU de Clermont-Ferrand, Clermont-Ferrand, France

Georgette, Berlier, Memory Resource and Research Centre of Saint-Etienne, CHU de Saint-Etienne, Saint-Etienne, France

Laëtitia, Berly, Memory Resource and Research Centre of Strasbourg, CHRU de Strasbourg, Strasbourg, France

Hassan, Berrissoul, Memory Resource and Research of Amiens, CHU Amiens Picardie, Amiens, France

Marie-Camille, Berthel, Memory Resource and Research Centre of Colmar, Colmar, France

Véronique, Berthier, Memory Resource and Research Centre of Lyon, Hospices Civils de Lyon, Lyon, France

François, Bertin-Hugault, Memory Resource and Research Centre of Lyon, Hospices Civils de Lyon, Lyon, France

François-Xavier, Bertrand, Memory Resource and Research Centre of Nantes, CHU de Nantes, Nantes, France

Guillaume, Bertrand, Memory Clinic of Avicenne, Hôpital Avicenne, Bobigny, France

Anaïck, Besozzi, Memory Resource and Research Centre of Nancy, CHU de Nancy, Nancy, France

Christelle, Betogliati-Filleau, Memory Resource and Research Centre of Nice, CHU de Nice, Nice, France

Catherine, Beze, Memory Resource and Research Centre of Tours, CHRU de Tours, Tours, France

Mathias, Bilger, Memory Resource and Research Centre of Strasbourg, CHRU de Strasbourg, Strasbourg, France

Sandrine, Bioux, Memory Resource and Research Centre of Rouen, CHU de Rouen, Rouen, France

Elisa, Bittard, Memory Resource and Research Centre of Bordeaux, CHU de Bordeaux, Bordeaux, France

Ludovic, Blanchard, Memory Resource and Research Centre of Poitiers, CHU de Poitiers, Poitiers, France

Odile, Blanchet, Memory Resource and Research Centre of Angers, CHU d'Angers, Angers, France

Maryline, Blanchon, Memory Resource and Research Centre of Marseille, CHU de Marseille, Marseille, France

Evangéline, Bliaux, Memory Resource and Research Centre of Rouen, CHU de Rouen, Rouen, France

Pierre, Bohn, Memory Resource and Research Centre of Rouen, CHU de Rouen, Rouen, France

Stéphanie, Bombois, Memory Resource and Research Centre of Lille, CHRU de Lille, Lille, France

Alain, Bonafe, Memory Resource and Research Centre of Montpellier, CHU de Montpellier, Montpellier, France

Marie, Bonnet, Memory Resource and Research Centre of Bordeaux, CHU de Bordeaux, Bordeaux, France

Hélène, Bonnot, Memory Resource and Research Centre of Strasbourg, CHRU de Strasbourg, Strasbourg, France

Martine, Bordessoules, Memory Resource and Research Centre of Bordeaux, CHU de Bordeaux, Bordeaux, France

Nathalie, Bortone, Memory Resource and Research Centre of Nice, CHU de Nice, Nice, France

Amandine, Bossant, Memory Resource and Research Centre of Grenoble, CHU de Grenoble Alpes,  Grenoble, France

Elodie, Bouaziz, Phar, Memory Resource and Research Centre of Paris Nord, AP-HP, Paris, France

Yasmina, Boudali, Memory Resource and Research Centre of Paris Broca, AP-HP, Paris, France

Laurie, Boukadida, Institute of Memory and Alzheimer's Disease (IM2A), France and Brain and Spine Institute (ICM), France UMR S 1127, Department of Neurology, AP-HP, Pitié-Salpêtrière University Hospital, Sorbonne Universities, Pierre et Marie Curie University, Paris, France

Justine, Boulanghien, Memory Resource and Research Centre of Montpellier, CHU de Montpellier, Montpellier, France

Clemence, Boully, Memory Resource and Research Centre of Paris Broca, AP-HP, Paris, France

Isabelle, Bourdel-Marchasson, Memory Resource and Research Centre of Bordeaux, CHU de Bordeaux, Bordeaux, France

Marie-France, Bourin, Memory Resource and Research Centre of Poitiers, CHU de Poitiers, Poitiers, France

Christophe, Bouvier, Coordinating Centre, CIC-1401 Clinical Epidemiology, Bordeaux, France

Serge, Bracard, Memory Resource and Research Centre of Nancy, CHU de Nancy, Nancy, France

Antoine, Brangier, Memory Resource and Research Centre of Angers, CHU d'Angers, Angers, France

Laëtitia, Breuilh, Memory Resource and Research Centre of Lille, CHRU de Lille, Lille, France

Lysiane, Brick, Memory Resource and Research Centre of Tours, CHRU de Tours, Tours, France

Marie-Laure, Brickert, Memory Resource and Research Centre of Colmar, Colmar, France

Pierre-Yves, Brillet, Memory Clinic of Avicenne, Hôpital Avicenne, Bobigny, France

Signe, Brinck, Memory Resource and Research Centre of Nice, CHU de Nice, Nice, France

Caroline, Buisson, Memory Resource and Research Centre of Bordeaux, CHU de Bordeaux, Bordeaux, France

Francine, Bury, Memory Resource and Research Centre of Colmar, Colmar, France

Laurence, Cadet, Memory Resource and Research Centre of Saint-Etienne, CHU de Saint-Etienne, Saint-Etienne, France

Julien, Cahors, Memory Resource and Research Centre of Nice, CHU de Nice, Nice, France

Laure, Caillard, Memory Resource and Research Centre of Paris Broca, AP-HP, Paris, France

Maria, Callejo, Plazas, Memory Resource and Research Centre of Nice, CHU de Nice, Nice, France

Fabienne, Calvas, Memory Resource and Research Centre of Toulouse, CHU de Toulouse, Toulouse, France

Sabine, Camara, Memory Resource and Research Centre of Colmar, Colmar, France

Aurore, Camoreyt, Memory Resource and Research Centre of Colmar, Colmar, France

Sandra, Campagne, Memory Resource and Research Centre of Marseille, CHU de Marseille, Marseille, France

Agnès, Camus, Memory Resource and Research Centre of Dijon, CHU Dijon Bourgogne, Dijon, France

Vincent, Camus, Memory Resource and Research Centre of Tours, CHRU de Tours, Tours,  France

Sandrine, Canaple, Memory Resource and Research of Amiens, CHU Amiens Picardie, Amiens, France

Edith, Carneiro, Memory Resource and Research Centre of Toulouse, CHU de Toulouse, Toulouse, France

Sabine, Caron, Memory Resource and Research Centre of Lille, CHRU de Lille, Lille, France

Antoine, Carpentier, Memory Clinic of Avicenne, Hôpital Avicenne, Bobigny, France

Elise, Carré, Memory Resource and Research Centre of Lille, CHRU de Lille, Lille, France

Isabelle, Carrie, Memory Resource and Research Centre of Toulouse, CHU de Toulouse, Toulouse, France

Pascaline, Cassagnaud, Memory Resource and Research Centre of Lille, CHRU de Lille, Lille, France

Françoise, Cattin, Memory Resource and Research Centre of Besançon, Besançon, France

Valérie, Causse-Lemercier, Laboratoire d'Imagerie Biomédicale, Sorbonne Universités, UPMC Univ Paris 06, Inserm U 1146, CNRS UMR 7371, F-75006 Paris, France NeuroSpin, I2BM, Commissariat à l'Energie Atomique, France

Anne, Cavey, Memory Resource and Research Centre of Nice, CHU de Nice, Nice, France

Matthieu, Chabel, Memory Resource and Research Centre of Lille, CHRU de Lille, Lille, France

Ludivine, Chamard, Memory Resource and Research Centre of Besançon, Besançon, France

Stéphane, Chanalet, Memory Resource and Research Centre of Nice, CHU de Nice, Nice, France

Thierry, Chaptal, Memory Resource and Research Centre of Montpellier, CHU de Montpellier, Montpellier, France

Annik, Charnallet, Memory Resource and Research Centre of Grenoble, CHU de Grenoble Alpes, Grenoble, France

Hélène, Chartrel, Memory Resource and Research Centre of Angers, CHU d'Angers, Angers, France

Mathieu, Chastan, Memory Resource and Research Centre of Rouen, CHU de Rouen, Rouen, France

Rose-May, Chaudat, Memory Resource and Research Centre of Marseille, CHU de Marseille, Marseille, France

Sophie, Chauvelier, Memory Resource and Research Centre of Paris Broca, AP-HP, Paris, France

Valérie, Chauvire, Memory Resource and Research Centre of Angers, CHU d'Angers, Angers, France

Samia, Cheriet, Memory Resource and Research Centre of Toulouse, CHU de Toulouse, Toulouse, France

Sylvie, Chiron, Memory Resource and Research Centre of Colmar, Colmar, France

Gilles, Chopard, Memory Resource and Research Centre of Besançon, Besançon, France

Emilie, Chrétien, Memory Resource and Research Centre of Lyon, Hospices Civils de Lyon, Lyon, France

Dominique, Clamens, Memory Resource and Research Centre of Montpellier, CHU de Montpellier, Montpellier, France

Anthony, Clotagatide, Memory Resource and Research Centre of Saint-Etienne, CHU de Saint-Etienne, Saint-Etienne, France

Emmanuel, Cognat, Memory Resource and Research Centre of Paris Nord, AP-HP, Paris, France

Lora, Cohen, Memory Resource and Research Centre of Grenoble, CHU de Grenoble Alpes,  Grenoble, France

Olivier, Colliot, Institute of Memory and Alzheimer's Disease (IM2A), France and Brain and Spine Institute (ICM), France UMR S 1127, Department of Neurology, AP-HP, Pitié-Salpêtrière University Hospital, Sorbonne Universities, Pierre et Marie Curie University, Paris, France

Jean-Marc, Constans, Memory Resource and Research of Amiens, CHU Amiens Picardie, Amiens, France

Elodie, Cordier, Memory Resource and Research Centre of Lille, CHRU de Lille, Lille, France

Marie-Hélène, Coste, Memory Resource and Research Centre of Lyon, Hospices Civils de Lyon, Lyon, France

Jean-Philippe, Cottier, Memory Resource and Research Centre of Tours, CHRU de Tours, Tours,  France

François, Cotton, Memory Resource and Research Centre of Lyon, Hospices Civils de Lyon, Lyon, France

Pierre, Malick, Coulibaly, Memory Resource and Research Centre of Nice, CHU de Nice, Nice, France

Isabelle, Couret, Memory Resource and Research Centre of Montpellier, CHU de Montpellier, Montpellier, France

Françoise, Courtin, Memory Resource and Research of Amiens, CHU Amiens Picardie, Amiens, France

Olivier-François, Couturier, Memory Resource and Research Centre of Angers, CHU d'Angers, Angers, France

Pascale, Cowppli-Bony, Memory Resource and Research Centre of Bordeaux, CHU de Bordeaux, Bordeaux, France

Véronique, Cressot, Memory Resource and Research Centre of Bordeaux, CHU de Bordeaux, Bordeaux, France

Benjamin, Crétin, Memory Resource and Research Centre of Strasbourg, CHRU de Strasbourg, Strasbourg, France

Marie-Hélène, Criscione, Memory Clinic of Avicenne, Hôpital Avicenne, Bobigny, France

Laurie, Cuche, Memory Resource and Research Centre of Saint-Etienne, CHU de Saint-Etienne, Saint-Etienne, France

Audrey, Dalbo, Memory Resource and Research Centre of Bordeaux, CHU de Bordeaux, Bordeaux, France

Keren, Danaila, Memory Resource and Research Centre of Lyon, Hospices Civils de Lyon, Lyon, France

Sabine, Dantzer, Memory Resource and Research Centre of Strasbourg, CHRU de Strasbourg, Strasbourg, France

Frédérique, Darcourt, Memory Resource and Research Centre of Nice, CHU de Nice, Nice, France

Jacques, Darcourt, Memory Resource and Research Centre of Nice, CHU de Nice, Nice, France

Pascal, Dartois, Memory Resource and Research Centre of Lille, CHRU de Lille, Lille, France

Ana-Maria, Dascalita, Memory Resource and Research Centre of Saint-Etienne, CHU de Saint-Etienne, Saint-Etienne, France

Marie-Claude, Daudon, Memory Resource and Research Centre of Saint-Etienne, CHU de Saint-Etienne, Saint-Etienne, France

Francesca, De, Anna, Memory Resource and Research Centre of Marseille, CHU de Marseille, Marseille, France

Virginie, de, Beco, Memory Clinic of Avicenne, Hôpital Avicenne, Bobigny, France

Xavier, de, Petigny, Memory Resource and Research Centre of Strasbourg, CHRU de Strasbourg, Strasbourg, France

Delphine, De, Verbizier-Lonjon, Memory Resource and Research Centre of Montpellier, CHU de Montpellier, Montpellier, France

Marielle, Decousus, Memory Resource and Research Centre of Saint-Etienne, CHU de Saint-Etienne, Saint-Etienne, France

Isabelle, Defouilloy, Memory Resource and Research of Amiens, CHU Amiens Picardie, Amiens, France

Cécile, Delaunay-Bretaut, Memory Resource and Research Centre of Angers, CHU d'Angers, Angers, France

Xavier, Delbeuck, Memory Resource and Research Centre of Lille, CHRU de Lille, Lille, France

Melissa, Delhommeau, Memory Resource and Research Centre of Nice, CHU de Nice, Nice, France

Christine, Delmaire, Memory Resource and Research Centre of Lille, CHRU de Lille, Lille, France

Floriane, Delphin-Combe, Memory Resource and Research Centre of Lyon, Hospices Civils de Lyon, Lyon, France

Julien, Delrieu, Memory Resource and Research Centre of Toulouse, CHU de Toulouse, Toulouse, France

Catherine, Demuyinck, Memory Resource and Research Centre of Strasbourg, CHRU de Strasbourg, Strasbourg, France

Vincent, Deramecourt, Memory Resource and Research Centre of Lille, CHRU de Lille, Lille, France

Hervé, Deramond, Memory Resource and Research of Amiens, CHU Amiens Picardie, Amiens, France

Virginie, Derenaucourt, Memory Resource and Research Centre of Lille, CHRU de Lille, Lille, France

Thomas, Desmidt, Memory Resource and Research Centre of Tours, CHRU de Tours, Tours,  France

Marie-Dominique, Desruet, Memory Resource and Research Centre of Grenoble, CHU de Grenoble Alpes,  Grenoble, France

Julien, Detour, Memory Resource and Research Centre of Strasbourg, CHRU de Strasbourg, Strasbourg, France

Audrey, Deudon, Memory Resource and Research Centre of Nice, CHU de Nice, Nice, France

Viviane, Derreux, Coordinating Centre, CIC-1401 Clinical Epidemiology, Administrative Assistant, France

Agnès, Devendeville, Memory Resource and Research of Amiens, CHU Amiens Picardie, Amiens, France

Laure, Di, Bitonto, Memory Resource and Research Centre of Strasbourg, CHRU de Strasbourg, Strasbourg, France

Sally, Dia, Memory Resource and Research Centre of Colmar, Colmar, France

Mira, Didic, Memory Resource and Research Centre of Marseille, CHU de Marseille, Marseille, France

Maritchu, Doireau, Memory Resource and Research Centre of Bordeaux, CHU de Bordeaux, Bordeaux, France

Marie-Thérèse, Dorier, Memory Resource and Research Centre of Besançon, Besançon, France

Antonio, Dos, Santos, Institute of Memory and Alzheimer's Disease (IM2A), France and Brain and Spine Institute (ICM), France UMR S 1127, Department of Neurology, AP-HP, Pitié-Salpêtrière University Hospital, Sorbonne Universities, Pierre et Marie Curie University, Paris, France

Patrice, Douillet, Memory Resource and Research Centre of Montpellier, CHU de Montpellier, Montpellier, France

Déborah, Drai, Memory Resource and Research Centre of Lyon, Hospices Civils de Lyon, Lyon, France

Foucaud, Du, Boisgueheneuc, Memory Resource and Research Centre of Poitiers, CHU de Poitiers, Poitiers, France

Delphine, Dubail, Memory Resource and Research Centre of Paris Broca, AP-HP, Paris, France

Sandrine, Duchez, Memory Resource and Research Centre of Saint-Etienne, CHU de Saint-Etienne, Saint-Etienne, France

Nathalie, Dufay, Memory Resource and Research Centre of Lyon, Hospices Civils de Lyon, Lyon, France

Sophie, Dulhoste, Memory Resource and Research Centre of Bordeaux, CHU de Bordeaux, Bordeaux, France

Julien, Dumont, Memory Resource and Research Centre of Lille, CHRU de Lille, Lille, France

Julien, Dumurgier, Memory Resource and Research Centre of Paris Nord, AP-HP, Paris, France

Mélanie, Dupin, Memory Resource and Research Centre of Saint-Etienne, CHU de Saint-Etienne, Saint-Etienne, France

Diane, Dupuy, Memory Resource and Research of Amiens, CHU Amiens Picardie, Amiens, France

Emmanuelle, Durand, Memory Resource and Research Centre of Bordeaux, CHU de Bordeaux, Bordeaux, France

Emmanuelle, Duron, Memory Resource and Research Centre of Paris Broca, AP-HP, Paris, France

Inna, Dygai-Cochet, Memory Resource and Research Centre of Dijon, CHU Dijon Bourgogne, Dijon, France

Véronique, Eder, Memory Clinic of Avicenne, Hôpital Avicenne, Bobigny, France

Emmanuelle, Ehrhard, Memory Resource and Research Centre of Strasbourg, CHRU de Strasbourg, Strasbourg, France

Hanane, El, Haouari, Memory Resource and Research Centre of Saint-Etienne, CHU de Saint-Etienne, Saint-Etienne, France

Elise, Enderlin, Memory Resource and Research Centre of Strasbourg, CHRU de Strasbourg, Strasbourg, France

Stéphane, Epelbaum, Institute of Memory and Alzheimer's Disease (IM2A), France and Brain and Spine Institute (ICM), France UMR S 1127, Department of Neurology, AP-HP, Pitié-Salpêtrière University Hospital, Sorbonne Universities, Pierre et Marie Curie University, Paris, France

Julie, Erraud, CIC-1401 Clinical Epidemiology, CHU de Bordeaux, Bordeaux, France

Frédérique, Etcharry-Bouyx, Memory Resource and Research Centre of Angers, CHU d'Angers, Angers, France

Magali, Eyriey, Memory Resource and Research Centre of Colmar, Colmar, France

Daniel, Fagret, Memory Resource and Research Centre of Grenoble, CHU de Grenoble Alpes,  Grenoble, France

Isabelle, Faillenot, Memory Resource and Research Centre of Saint-Etienne, CHU de Saint-Etienne, Saint-Etienne, France

Catherine, Faisant, Memory Resource and Research Centre of Toulouse, CHU de Toulouse, Toulouse, France

Karim, Farid, Memory Resource and Research Centre of Paris Nord, AP-HP, Paris, France

Véronique, Fasquel, Memory Resource and Research of Amiens, CHU Amiens Picardie, Amiens, France

Marion, Fatisson, Memory Resource and Research Centre of Saint-Etienne, CHU de Saint-Etienne, Saint-Etienne, France

Denis, Fédérico, Memory Resource and Research Centre of Lyon, Hospices Civils de Lyon, Lyon, France

Olivier, Felician, Memory Resource and Research Centre of Marseille, CHU de Marseille, Marseille, France

Philippe, Fernandez, Memory Resource and Research Centre of Bordeaux, CHU de Bordeaux, Bordeaux, France

Sabrina, Ferreira, Memory Resource and Research Centre of Besançon, Besançon, France

Camille, Ferté, Memory Resource and Research Centre of Lille, CHRU de Lille, Lille, France

Guillaume, Fiard, Memory Resource and Research Centre of Lyon, Hospices Civils de Lyon, Lyon, France

Florine, Fievet, Memory Resource and Research Centre of Lille, CHRU de Lille, Lille, France

Martine, Flores, Memory Resource and Research Centre of Montpellier, CHU de Montpellier, Montpellier, France

Pacôme, Fosse, Memory Resource and Research Centre of Angers, CHU d'Angers, Angers, France

Alexandra, Foubert-Samier, Memory Resource and Research Centre of Bordeaux, CHU de Bordeaux, Bordeaux, France

Sandrine, Fouchet, Memory Resource and Research Centre of Bordeaux, CHU de Bordeaux, Bordeaux, France

Marjolaine, Fourcade, Memory Resource and Research Centre of Montpellier, CHU de Montpellier, Montpellier, France

Isabelle, Franck, Memory Resource and Research Centre of Strasbourg, CHRU de Strasbourg, Strasbourg, France

Monique, Galitzky, Memory Resource and Research Centre of Toulouse, CHU de Toulouse, Toulouse, France

Céline, Gallazzini-Crepin, Memory Resource and Research Centre of Grenoble, CHU de Grenoble Alpes,  Grenoble, France

Radka, Gantcheva, Memory Resource and Research Centre of Marseille, CHU de Marseille, Marseille, France

Laurence, Garbarg-Chenon, Memory Clinic of Avicenne, Hôpital Avicenne, Bobigny, France

Patrick, Gelé, Memory Resource and Research Centre of Lille, CHRU de Lille, Lille, France

Emmanuel, Gerardin, Memory Resource and Research Centre of Rouen, CHU de Rouen, Rouen, France

Pascale, Gerardin, Memory Resource and Research Centre of Nancy, CHU de Nancy, Nancy, France

Loïc, Gerlier, Memory Resource and Research Centre of Bordeaux, CHU de Bordeaux, Bordeaux, France

Claire, Gervais, Memory Resource and Research Centre of Nice, CHU de Nice, Nice, France

Jean-Claude, Getenet, Memory Resource and Research Centre of Saint-Etienne, CHU de Saint-Etienne, Saint-Etienne, France

Cindy, Giaume, Memory Resource and Research Centre of Nice, CHU de Nice, Nice, France

Carole, Girard, Memory Resource and Research Centre of Rouen, CHU de Rouen, Rouen, France

Nadine, Girard, Memory Resource and Research Centre of Marseille, CHU de Marseille, Marseille, France

Béatrice, Giroz, Memory Resource and Research Centre of Strasbourg, CHRU de Strasbourg, Strasbourg, France

Chantal, Girtanner, Memory Resource and Research Centre of Saint-Etienne, CHU de Saint-Etienne, Saint-Etienne, France

Valérie, Gissot, Memory Resource and Research Centre of Tours, CHRU de Tours, Tours, France

Blandine, Giusti, Memory Resource and Research Centre of Lyon, Hospices Civils de Lyon, Lyon, France

Patrick, Gouel, Memory Resource and Research Centre of Rouen, CHU de Rouen, Rouen, France

Natalina, Gour, Memory Resource and Research Centre of Marseille, CHU de Marseille, Marseille, France

Anne-Sophie, Gourgues, Memory Resource and Research Centre of Poitiers, CHU de Poitiers, Poitiers, France

Caroline, Grangeon, Memory Resource and Research Centre of Nice, CHU de Nice, Nice, France

Caroline, Grasselli-Monboisse, Memory Resource and Research Centre of Montpellier, CHU de Montpellier, Montpellier, France

Hélène, Gros-Dagnac, Memory Resource and Research Centre of Toulouse, CHU de Toulouse, Toulouse, France

Daniel, Grucker, Memory Resource and Research Centre of Strasbourg, CHRU de Strasbourg, Strasbourg, France

Eric, Guedj, Memory Resource and Research Centre of Marseille, CHU de Marseille, Marseille, France

Claude, Gueriot, Memory Resource and Research Centre of Marseille, CHU de Marseille, Marseille, France

Blandine, Guignard, Memory Resource and Research Centre of Strasbourg, CHRU de Strasbourg, Strasbourg, France

Yves, Guilhermet, Memory Resource and Research Centre of Lyon, Hospices Civils de Lyon, Lyon, France

Rémy, Guillevin, Memory Resource and Research Centre of Poitiers, CHU de Poitiers, Poitiers, France

Anne, Guyard, Memory Resource and Research Centre of Angers, CHU d'Angers, Angers, France

Jacques, Guyard, Memory Resource and Research Centre of Angers, CHU d'Angers, Angers, France

Lilia, Habbessi, Memory Resource and Research Centre of Lyon, Hospices Civils de Lyon, Lyon, France

Sophie, Haffen, Memory Resource and Research Centre of Besançon, Besançon, France

Sarah, Hammami, Memory Clinic of Avicenne, Hôpital Avicenne, Bobigny, France

Didier, Hannequin, Memory Resource and Research Centre of Rouen, CHU de Rouen, Rouen, France

Véronique, Hannier, Memory Resource and Research Centre of Rouen, CHU de Rouen, Rouen, France

Anne-Marie, Hanser, Memory Resource and Research Centre of Colmar, Colmar, France

Saoussen, Haouas, Memory Resource and Research Centre of Paris Broca, AP-HP, Paris, France

Anaïs, Heurtebise, Memory Resource and Research Centre of Montpellier, CHU de Montpellier, Montpellier, France

Sophie, Hierry, Memory Resource and Research Centre of Colmar, Colmar, France

Anne, Hitzel, Memory Resource and Research Centre of Toulouse, CHU de Toulouse, Toulouse, France

Claude, Hossein-Foucher, Memory Resource and Research Centre of Lille, CHRU de Lille, Lille, France

Fabrice, Hubele, Memory Resource and Research Centre of Strasbourg, CHRU de Strasbourg, Strasbourg, France

Sabrina, Iannuzzi, Memory Resource and Research Centre of Grenoble, CHU de Grenoble Alpes, Grenoble, France

Danielle, Ibarrola, Memory Resource and Research Centre of Lyon, Hospices Civils de Lyon, Lyon, France

Sandrine, Indart, Memory Resource and Research Centre of Paris Nord, AP-HP, Paris, France

Agnès, Jacquin-Piques, Memory Resource and Research Centre of Dijon, CHU Dijon Bourgogne, Dijon, France

Sophie, Jaeger, Memory Resource and Research Centre of Colmar, Colmar, France

Séverine, Jallier, Coordinating Centre, CIC-1401 Clinical Epidemiology, Bordeaux, France

Betty, Jean, Memory Resource and Research Centre of Clermont-Ferrand, CHU de Clermont-Ferrand, Clermont-Ferrand, France

Joanne, Jenn, Memory Resource and Research Centre of Bordeaux, CHU de Bordeaux, Bordeaux, France

Laure, Joly, Memory Resource and Research Centre of Nancy, CHU de Nancy, Nancy, France

Thérèse, Jonveaux, Memory Resource and Research Centre of Nancy, CHU de Nancy, Nancy, France

Séverine, Jourdain, Memory Resource and Research Centre of Rouen, CHU de Rouen, Rouen, France

Adrien, Julian, Memory Resource and Research Centre of Poitiers, CHU de Poitiers, Poitiers, France

Barbara, Jung, Memory Resource and Research Centre of Strasbourg, CHRU de Strasbourg, Strasbourg, France

Alexandra, Juphard, Memory Resource and Research Centre of Grenoble, CHU de Grenoble Alpes,  Grenoble, France

Nora, Karaoun, Memory Resource and Research Centre of Paris Nord, AP-HP, Paris, France

Anisse, Karoun, CIC-1401 Clinical Epidemiology, CHU de Bordeaux, Bordeaux, France

Aurélie, Kas, Laboratoire d'Imagerie Biomédicale, Sorbonne Universités, UPMC Univ Paris 06, Inserm U 1146, CNRS UMR 7371, F-75006 Paris, France NeuroSpin, I2BM, Commissariat à l'Energie Atomique, France

Anna, Kearney-Schwartz, Memory Resource and Research Centre of Nancy, CHU de Nancy, Nancy, France

Sandrine, Keignart, Memory Resource and Research Centre of Grenoble, CHU de Grenoble Alpes,  Grenoble, France

Antony, Kelly, Memory Resource and Research Centre of Clermont-Ferrand, CHU de Clermont-Ferrand, Clermont-Ferrand, France

Anne, Klebaur, Memory Resource and Research Centre of Colmar, Colmar, France

Catherine, Kleitz, Memory Resource and Research Centre of Strasbourg, CHRU de Strasbourg, Strasbourg, France

Lejla, Koric, Memory Resource and Research Centre of Marseille, CHU de Marseille, Marseille, France

Alexandre, Krainik, Memory Resource and Research Centre of Grenoble, CHU de Grenoble Alpes,  Grenoble, France

Stéphane, Kremer, Memory Resource and Research Centre of Strasbourg, CHRU de Strasbourg, Strasbourg, France

Florian, Labourée, Memory Resource and Research Centre of Paris Broca, AP-HP, Paris, France

Franck, Lacoeuille, Memory Resource and Research Centre of Angers, CHU d'Angers, Angers, France

Valérie, Lafont, Memory Resource and Research Centre of Nice, CHU de Nice, Nice, France

Marie-Claude, Lagneau, Memory Resource and Research Centre of Dijon, CHU Dijon Bourgogne, Dijon, France

Sophie, Lagouarde, Memory Resource and Research Centre of Bordeaux, CHU de Bordeaux, Bordeaux, France

Francoise, Lala, Memory Resource and Research Centre of Toulouse, CHU de Toulouse, Toulouse, France

Frédéric, Lamare, Memory Resource and Research Centre of Bordeaux, CHU de Bordeaux, Bordeaux, France

Sophie, Lamarque, CIC-1401 Clinical Epidemiology, CHU de Bordeaux, Bordeaux, France

Franck, Lamberton, Memory Resource and Research Centre of Lyon, Hospices Civils de Lyon, Lyon, France

Chantal, Lamy, Memory Resource and Research of Amiens, CHU Amiens Picardie, Amiens, France

Pauline, Lapalus, Memory Resource and Research Centre of Paris Nord, AP-HP, Paris, France

Jean-Louis, Laplanche, Memory Resource and Research Centre of Paris Nord, AP-HP, Paris, France

Delphine, Lassus-Sangosse, Memory Resource and Research Centre of Grenoble, CHU de Grenoble Alpes,  Grenoble, France

Caroline, Latger-Florence, Memory Resource and Research Centre of Marseille, CHU de Marseille, Marseille, France

Cyrille, Launay, Memory Resource and Research Centre of Angers, CHU d'Angers, Angers, France

Caroline, Laurent, Memory Resource and Research Centre of Lyon, Hospices Civils de Lyon, Lyon, France

Mathilde, Laye, Memory Resource and Research Centre of Nice, CHU de Nice, Nice, France

Didier, Le, Bars, Memory Resource and Research Centre of Lyon, Hospices Civils de Lyon, Lyon, France

Séverine, Le, Dily, Memory Resource and Research Centre of Nantes, CHU de Nantes, Nantes, France

Liliane, Le, Guay, Memory Resource and Research Centre of Strasbourg, CHRU de Strasbourg, Strasbourg, France

Lisa, Le, Scouarnec, CIC-1401 Clinical Epidemiology, CHU de Bordeaux, Bordeaux, France

Isabelle, Le, Taillandier, de, Gabory, Memory Resource and Research Centre of Bordeaux, CHU de Bordeaux, Bordeaux, France

Emmanuelle, Lebars, Memory Resource and Research Centre of Montpellier, CHU de Montpellier, Montpellier, France

Cécile, Lebrun-Givois, Memory Resource and Research Centre of Saint-Etienne, CHU de Saint-Etienne, Saint-Etienne, France

Eugénie, Leclerc, Memory Clinic of Avicenne, Hôpital Avicenne, Bobigny, France

Jihyun, Lee, Roy, Memory Resource and Research Centre of Nice, CHU de Nice, Nice, France

Jean-François, Legrand, Memory Resource and Research Centre of Lille, CHRU de Lille, Lille, France

Stéphane, Lehericy, Institute of Memory and Alzheimer's Disease (IM2A), France and Brain and Spine Institute (ICM), France UMR S 1127, Department of Neurology, AP-HP, Pitié-Salpêtrière University Hospital, Sorbonne Universities, Pierre et Marie Curie University, Paris, France

Sylvain, Lehmann, Memory Resource and Research Centre of Montpellier, CHU de Montpellier, Montpellier, France

Mathieu, Leininger, Memory Resource and Research Centre of Nancy, CHU de Nancy, Nancy, France

Justine, Lemaire, Memory Resource and Research Centre of Nice, CHU de Nice, Nice, France

Hermine, Lenoir, Memory Resource and Research Centre of Paris Broca, AP-HP, Paris, France

Marylin, Leny, Memory Resource and Research Centre of Paris Nord, AP-HP, Paris, France

Elsa, Leone, Memory Resource and Research Centre of Nice, CHU de Nice, Nice, France

Mélanie, Leroy, Memory Resource and Research Centre of Lille, CHRU de Lille, Lille, France

Mylène, Lesage, Memory Resource and Research Centre of Strasbourg, CHRU de Strasbourg, Strasbourg, France

Marcel, Levy, Institute of Memory and Alzheimer's Disease (IM2A), France and Brain and Spine Institute (ICM), France UMR S 1127, Department of Neurology, AP-HP, Pitié-Salpêtrière University Hospital, Sorbonne Universities, Pierre et Marie Curie University, Paris, France

Stéphanie, Libercier, Memory Resource and Research Centre of Colmar, Colmar, France

Julie, Lidier, CIC-1401 Clinical Epidemiology, CHU de Bordeaux, Bordeaux, France

Nadine, Longato, Memory Resource and Research Centre of Strasbourg, CHRU de Strasbourg, Strasbourg, France

Paulo, Loureiro, de, Sousa, Memory Resource and Research Centre of Strasbourg, CHRU de Strasbourg, Strasbourg, France

Marie, Luce, Royère, Memory Resource and Research Centre of Marseille, CHU de Marseille, Marseille, France

Juliette, Ly, Institute of Memory and Alzheimer's Disease (IM2A), France and Brain and Spine Institute (ICM), France UMR S 1127, Department of Neurology, AP-HP, Pitié-Salpêtrière University Hospital, Sorbonne Universities, Pierre et Marie Curie University, Paris, France

Marie-Anne, Mackowiak-Cordoliani, Memory Resource and Research Centre of Lille, CHRU de Lille, Lille, France

Eloi, Magnin, Memory Resource and Research Centre of Besançon, Besançon, France

Serge, Maia, Memory Resource and Research Centre of Tours, CHRU de Tours, Tours, France

Didier, Maillet, Memory Clinic of Avicenne, Hôpital Avicenne, Bobigny, France

Zaza, Makaroff, Memory Resource and Research Centre of Lyon, Hospices Civils de Lyon, Lyon, France

Oldès, Mansour, Memory Resource and Research Centre of Lyon, Hospices Civils de Lyon, Lyon, France

Athina, Marantidou, Memory Clinic of Avicenne, Hôpital Avicenne, Bobigny, France

Isabelle, Marcet, Memory Resource and Research Centre of Bordeaux, CHU de Bordeaux, Bordeaux, France

Olivier, Marcy, CIC-1401 Clinical Epidemiology, CHU de Bordeaux, Bordeaux, France

Cécilia, Marelli, Memory Resource and Research Centre of Montpellier, CHU de Montpellier, Montpellier, France

Sophie, Marilier, Memory Resource and Research Centre of Dijon, CHU Dijon Bourgogne, Dijon, France

Fanny, Marmet, Memory Resource and Research Centre of Nice, CHU de Nice, Nice, France

Laurent, Marquine, Memory Resource and Research Centre of Toulouse, CHU de Toulouse, Toulouse, France

Corinne, Marrer, Memory Resource and Research Centre of Strasbourg, CHRU de Strasbourg, Strasbourg, France

Idalie, Martin, Memory Resource and Research Centre of Lyon, Hospices Civils de Lyon, Lyon, France

Sandrine, Martin, Memory Resource and Research Centre of Montpellier, CHU de Montpellier, Montpellier, France

Olivier, Martinaud, Memory Resource and Research Centre of Rouen, CHU de Rouen, Rouen, France

Catherine, Martin-Hunyadi, Memory Resource and Research Centre of Strasbourg, CHRU de Strasbourg, Strasbourg, France

Isabelle, Mathieu, Memory Resource and Research Centre of Colmar, Colmar, France

Fabien, Maurel, Memory Resource and Research Centre of Nice, CHU de Nice, Nice, France

Sylvie, Maymon, Memory Resource and Research Centre of Colmar, Colmar, France

Joachim, Mazère, Memory Resource and Research Centre of Bordeaux, CHU de Bordeaux, Bordeaux, France

Aïcha, Medjoul, Memory Clinic of Avicenne, Hôpital Avicenne, Bobigny, France

Isabelle, Meiss, Memory Resource and Research Centre of Strasbourg, CHRU de Strasbourg, Strasbourg, France

Aurélie, Méozoone, Memory Resource and Research Centre of Paris Nord, AP-HP, Paris, France

Isabelle, Merlet, Memory Resource and Research Centre of Poitiers, CHU de Poitiers, Poitiers, France

Catherine, Mertz, Memory Resource and Research Centre of Besançon, Besançon, France

Danielle, Mestas, Memory Resource and Research Centre of Clermont-Ferrand, CHU de Clermont-Ferrand, Clermont-Ferrand, France

Catherine, Metzger, Memory Resource and Research Centre of Strasbourg, CHRU de Strasbourg, Strasbourg, France

Sabine, Meurrens, Memory Resource and Research Centre of Lille, CHRU de Lille, Lille, France

Marc-Etienne, Meyer, Memory Resource and Research of Amiens, CHU Amiens Picardie, Amiens, France

Jean-Marc, Michel, Memory Resource and Research Centre of Colmar, Colmar, France

Agnès, Michon, Institute of Memory and Alzheimer's Disease (IM2A), France and Brain and Spine Institute (ICM), France UMR S 1127, Department of Neurology, AP-HP, Pitié-Salpêtrière University Hospital, Sorbonne Universities, Pierre et Marie Curie University, Paris, France

Isabelle, Migeon-Duballet, Memory Resource and Research Centre of Poitiers, CHU de Poitiers, Poitiers, France

Carole, Miguet-Alfonsi, Memory Resource and Research Centre of Besançon, Besançon, France

Karl, Mondon, Memory Resource and Research Centre of Tours, CHRU de Tours, Tours, France

Laëtitia, Monjoin, Memory Resource and Research Centre of Strasbourg, CHRU de Strasbourg, Strasbourg, France

Pascale, Morel, Memory Resource and Research Centre of Colmar, Colmar, France

Sébastien, Moreno, Memory Resource and Research Centre of Nice, CHU de Nice, Nice, France

Clément, Morgat, Memory Resource and Research Centre of Bordeaux, CHU de Bordeaux, Bordeaux, France

Charline, Morillon, Memory Resource and Research Centre of Tours, CHRU de Tours, Tours, France

Chrystèle, Mosca, Memory Resource and Research Centre of Grenoble, CHU de Grenoble Alpes, Grenoble, France

Véronique, Moullart, Memory Resource and Research of Amiens, CHU Amiens Picardie, Amiens, France

Christian, Moussard, Memory Resource and Research Centre of Besançon, Besançon, France

Aurélie, Mouton, Memory Resource and Research Centre of Nice, CHU de Nice, Nice, France

Izzie, Jacques, Namer, Memory Resource and Research Centre of Strasbourg, CHRU de Strasbourg, Strasbourg, France

Jungalee, Navichka, Laboratoire d'Imagerie Biomédicale, Sorbonne Universités, UPMC Univ Paris 06, Inserm U 1146, CNRS UMR 7371, F-75006 Paris, France NeuroSpin, I2BM, Commissariat à l'Energie Atomique, France

Sophie, Navucet, Memory Resource and Research Centre of Montpellier, CHU de Montpellier, Montpellier, France

Raymond, Nelly, Memory Resource and Research Centre of Lyon, Hospices Civils de Lyon, Lyon, France

Thierry, Nicolas, Memory Resource and Research Centre of Besançon, Besançon, France

Georges, Niewiadomski, Memory Resource and Research Centre of Nice, CHU de Nice, Nice, France

Guillaume, Nivaggoni, Memory Resource and Research Centre of Nice, CHU de Nice, Nice, France

Marie, Noblet, Memory Resource and Research Centre of Strasbourg, CHRU de Strasbourg, Strasbourg, France

Nicolas, Noiret, Memory Resource and Research Centre of Besançon, Besançon, France

Fati, Nourhashemi, Memory Resource and Research Centre of Toulouse, CHU de Toulouse, Toulouse, France

Francis, Nyasse, Institute of Memory and Alzheimer's Disease (IM2A), France and Brain and Spine Institute (ICM), France UMR S 1127, Department of Neurology, AP-HP, Pitié-Salpêtrière University Hospital, Sorbonne Universities, Pierre et Marie Curie University, Paris, France

Estelle, Occelli, Memory Resource and Research Centre of Nice, CHU de Nice, Nice, France

Hélène, Oesterle, Memory Resource and Research Centre of Colmar, Colmar, France

Justine, Oosterlinck, Memory Resource and Research Centre of Lille, CHRU de Lille, Lille, France

Claudie, Ornon, Memory Resource and Research Centre of Poitiers, CHU de Poitiers, Poitiers, France

Galdric, Orvoen, Memory Resource and Research Centre of Paris Broca, AP-HP, Paris, France

Pierre, Jean, Ousset, Memory Resource and Research Centre of Toulouse, CHU de Toulouse, Toulouse, France

Anne, Pachart, Memory Resource and Research Centre of Colmar, Colmar, France

Florian, Palabaud, Memory Resource and Research Centre of Saint-Etienne, CHU de Saint-Etienne, Saint-Etienne, France

Juliette, Palisson, Memory Clinic of Avicenne, Hôpital Avicenne, Bobigny, France

Amandine, Pallardy, Memory Resource and Research Centre of Nantes, CHU de Nantes, Nantes, France

Sylvie, Papacatzis, Memory Resource and Research Centre of Grenoble, CHU de Grenoble Alpes,  Grenoble, France

Claire, Paquet, Memory Resource and Research Centre of Paris Nord, AP-HP, Paris, France

Pierre-Yves, Pare, Memory Resource and Research Centre of Angers, CHU d'Angers, Angers, France

Guillaume, Pariscoat, Memory Resource and Research Centre of Montpellier, CHU de Montpellier, Montpellier, France

Anne, Pasco, Memory Resource and Research Centre of Angers, CHU d'Angers, Angers, France

Pierre, Payoux, Memory Resource and Research Centre of Toulouse, CHU de Toulouse, Toulouse, France

Cécile, Pays, Memory Resource and Research Centre of Montpellier, CHU de Montpellier, Montpellier, France

Julie, Pelat, Memory Resource and Research Centre of Marseille, CHU de Marseille, Marseille, France

Katell, Peoch, Phar, Memory Resource and Research Centre of Paris Nord, AP-HP, Paris, France

Rémy, Perdrisot, Memory Resource and Research Centre of Poitiers, CHU de Poitiers, Poitiers, France

Raphaël, Pereira, Memory Resource and Research Centre of Lyon, Hospices Civils de Lyon, Lyon, France

Bertille, Perin, Memory Resource and Research of Amiens, CHU Amiens Picardie, Amiens, France

Christine, Perret-Guillaume, Memory Resource and Research Centre of Nancy, CHU de Nancy, Nancy, France

Sophie, Pérusat, Coordinating Centre, CIC-1401 Clinical Epidemiology, Clinical Project Manager, France

Yolande, Petegnief, Memory Resource and Research Centre of Besançon, Besançon, France

Grégory, Petyt, Memory Resource and Research Centre of Lille, CHRU de Lille, Lille, France

Lorène, Philibert, Memory Resource and Research Centre of Nice, CHU de Nice, Nice, France

Nathalie, Philippi, Memory Resource and Research Centre of Strasbourg, CHRU de Strasbourg, Strasbourg, France

Clélie, Phillipps, Memory Resource and Research Centre of Strasbourg, CHRU de Strasbourg, Strasbourg, France

Julie, Piano, Memory Resource and Research Centre of Nice, CHU de Nice, Nice, France

Michèle, Pierre, Memory Resource and Research Centre of Toulouse, CHU de Toulouse, Toulouse, France

Johan, Pietras, Memory Resource and Research Centre of Grenoble, CHU de Grenoble Alpes, Grenoble, France

Mélanie, Pigot, Memory Resource and Research Centre of Montpellier, CHU de Montpellier, Montpellier, France

Fanny, Pineau, Memory Resource and Research Centre of Colmar, Colmar, France

Geneviève, Pinganaud, Memory Resource and Research Centre of Bordeaux, CHU de Bordeaux, Bordeaux, France

Pierre, Pitet, Memory Resource and Research Centre of Grenoble, CHU de Grenoble Alpes, Grenoble, France

Matthieu, Plichart, Memory Resource and Research Centre of Paris Broca, AP-HP, Paris, France

Catherine, Poisson, Institute of Memory and Alzheimer's Disease (IM2A), France and Brain and Spine Institute (ICM), France UMR S 1127, Department of Neurology, AP-HP, Pitié-Salpêtrière University Hospital, Sorbonne Universities, Pierre et Marie Curie University, Paris, France

Elodie, Pongan, Memory Resource and Research Centre of Lyon, Hospices Civils de Lyon, Lyon, France

Gabriel, Pop, Memory Clinic of Avicenne, Hôpital Avicenne, Bobigny, France

Dorothée, Pouliquen, Memory Resource and Research Centre of Rouen, CHU de Rouen, Rouen, France

Cyril, Poupon, Institute of Memory and Alzheimer's Disease (IM2A), France and Brain and Spine Institute (ICM), France UMR S 1127, Department of Neurology, AP-HP, Pitié-Salpêtrière University Hospital, Sorbonne Universities, Pierre et Marie Curie University, Paris, France

Stéphane, Pouponneau, Memory Resource and Research Centre of Tours, CHRU de Tours, Tours, France

Bruno, Pozetto, Memory Resource and Research Centre of Saint-Etienne, CHU de Saint-Etienne, Saint-Etienne, France

Sophie, Pradier, Memory Resource and Research Centre of Bordeaux, CHU de Bordeaux, Bordeaux, France

Thierry, Prangère, Memory Resource and Research Centre of Lille, CHRU de Lille, Lille, France

Magali, Prévot, Memory Resource and Research Centre of Paris Nord, AP-HP, Paris, France

Evelyne, Provost, Memory Resource and Research Centre of Saint-Etienne, CHU de Saint-Etienne, Saint-Etienne, France

Michèle, Puel, Memory Resource and Research Centre of Toulouse, CHU de Toulouse, Toulouse, France

Mathieu, Queneau, Memory Resource and Research Centre of Paris Nord, AP-HP, Paris, France

Muriel, Quillard-Muraine, Memory Resource and Research Centre of Rouen, CHU de Rouen, Rouen, France

Valérie, Quipourt, Memory Resource and Research Centre of Dijon, CHU Dijon Bourgogne, Dijon, France

Chloé, Rachez, Memory Resource and Research Centre of Clermont-Ferrand, CHU de Clermont-Ferrand, Clermont-Ferrand, France

Aline, Rahnema, Memory Resource and Research Centre of Nancy, CHU de Nancy, Nancy, France

Muriel, Rainfray, Memory Resource and Research Centre of Bordeaux, CHU de Bordeaux, Bordeaux, France

Nadine, Raoux, Memory Resource and Research Centre of Bordeaux, CHU de Bordeaux, Bordeaux, France

Anatta, Razafimanantsoa, Memory Clinic of Avicenne, Hôpital Avicenne, Bobigny, France

Micheline, Razzouk-Cadet, Memory Resource and Research Centre of Nice, CHU de Nice, Nice, France

Maria, Rego-Lopes, Memory Resource and Research Centre of Paris Broca, AP-HP, Paris, France

Solveig, Relland, Memory Resource and Research Centre of Lyon, Hospices Civils de Lyon, Lyon, France

Marie, Revillon, Institute of Memory and Alzheimer's Disease (IM2A, France and Brain and Spine Institute (ICM, France UMR S 1127, Department of Neurology, AP-HP, Pitié-Salpêtrière University Hospital, Sorbonne Universities, Pierre et Marie Curie University, France

Sylvie, Richard, Memory Resource and Research Centre of Lyon, Hospices Civils de Lyon, Lyon, France

Virginie, Richard, Coordinating Centre, CIC-1401 Clinical Epidemiology, Data Manager, France

Eliane, Riera, Memory Resource and Research Centre of Colmar, Colmar, France

Anne-Sophie, Rigaud, Memory Resource and Research Centre of Paris Broca, AP-HP, Paris, France

Marie-Claire, Riocreux, Memory Resource and Research Centre of Saint-Etienne, CHU de Saint-Etienne, Saint-Etienne, France

Philippe, Robert, Memory Resource and Research Centre of Nice, CHU de Nice, Nice, France

Hélène, Robin-Ismer, Memory Resource and Research Centre of Strasbourg, CHRU de Strasbourg, Strasbourg, France

Laëtitia, Rocher, Memory Resource and Research Centre of Nantes, CHU de Nantes, Nantes, France

Fabienne, Rochette, Memory Resource and Research Centre of Grenoble, CHU de Grenoble Alpes,  Grenoble, France

Mathieu, Rodallec, Memory Resource and Research Centre of Paris Nord, AP-HP, Paris, France

Yves, Rolland, Memory Resource and Research Centre of Toulouse, CHU de Toulouse, Toulouse, France

Adeline, Rollin-Sillaire, Memory Resource and Research Centre of Lille, CHRU de Lille, Lille, France

Fabien, Rondepierre, Memory Resource and Research Centre of Clermont-Ferrand, CHU de Clermont-Ferrand, Clermont-Ferrand, France

Stéphanie, Roseng, Coordinating Centre, CIC-1401 Clinical Epidemiology, Clinical Research Associate, France

Mélanie, Rossitto, Memory Resource and Research Centre of Nancy, CHU de Nancy, Nancy, France

Caroline, Roubaud, Memory Resource and Research Centre of Lyon, Hospices Civils de Lyon, Lyon, France

Isabelle, Rouch, Memory Resource and Research Centre of Lyon, Hospices Civils de Lyon, Lyon, France

Olivier, Roulant, Memory Resource and Research Centre of Toulouse, CHU de Toulouse, Toulouse, France

Martine, Roussel, Memory Resource and Research of Amiens, CHU Amiens Picardie, Amiens, France

Annie, Routier, Memory Resource and Research of Amiens, CHU Amiens Picardie, Amiens, France

Julie, Roux, Memory Resource and Research Centre of Grenoble, CHU de Grenoble Alpes,  Grenoble, France

Perrine, Roy, Institute of Memory and Alzheimer's Disease (IM2A, France and Brain and Spine Institute (ICM, France UMR S 1127, Department of Neurology, AP-HP, Pitié-Salpêtrière University Hospital, Sorbonne Universities, Pierre et Marie Curie University, France

Séverine, Roy, Memory Resource and Research Centre of Lyon, Hospices Civils de Lyon, Lyon, France

Ilham, Ryff, Memory Resource and Research Centre of Besançon, Besançon, France

Guillaume, Sacco, Memory Resource and Research Centre of Nice, CHU de Nice, Nice, France

Djamel, Saidi, Memory Resource and Research Centre of Bordeaux, CHU de Bordeaux, Bordeaux, France

Anne-Sophie, Salabert, Memory Resource and Research Centre of Toulouse, CHU de Toulouse, Toulouse, France

François, Salmon, Memory Resource and Research Centre of Poitiers, CHU de Poitiers, Poitiers, France

Maria-Joao, Santiago-Ribeiro, Memory Resource and Research Centre of Tours, CHRU de Tours, Tours,  France

Joëlle, Sapin, Memory Resource and Research Centre of Bordeaux, CHU de Bordeaux, Bordeaux, France

Nadine, Sapin, Memory Resource and Research Centre of Nice, CHU de Nice, Nice, France

Alain, Sarciron, Memory Resource and Research Centre of Lyon, Hospices Civils de Lyon, Lyon, France

Nathalie, Sastres, Memory Resource and Research Centre of Toulouse, CHU de Toulouse, Toulouse, France

Amandine, Saubion, Memory Resource and Research Centre of Toulouse, CHU de Toulouse, Toulouse, France

Mathilde, Sauvée, Memory Resource and Research Centre of Grenoble, CHU de Grenoble Alpes,  Grenoble, France

Sophie, Schahl, Memory Resource and Research Centre of Colmar, Colmar, France

Christian, Scheiber, Memory Resource and Research Centre of Lyon, Hospices Civils de Lyon, Lyon, France

Aude, Schlecht, Memory Resource and Research Centre of Colmar, Colmar, France

Anne-Marie, Schneider, Memory Resource and Research Centre of Strasbourg, CHRU de Strasbourg, Strasbourg, France

Floraly, Sejalon, Memory Clinic of Avicenne, Hôpital Avicenne, Bobigny, France

Christiane, Sergent, Memory Resource and Research Centre of Bordeaux, CHU de Bordeaux, Bordeaux, France

Amélie, Serra, Memory Resource and Research Centre of Grenoble, CHU de Grenoble Alpes,  Grenoble, France

Marie-Laure, Seux, Memory Resource and Research Centre of Paris Broca, AP-HP, Paris, France

Romain, Simon, Memory Resource and Research Centre of Angers, CHU d'Angers, Angers, France

Valérie, Simon, Institute of Memory and Alzheimer's Disease (IM2A, France and Brain and Spine Institute (ICM, France UMR S 1127, Department of Neurology, AP-HP, Pitié-Salpêtrière University Hospital, Sorbonne Universities, Pierre et Marie Curie University, France

Rémi, Sitta, Coordinating Centre, CIC-1401 Clinical Epidemiology, France

Hélène, Sordet-Guépet, Memory Resource and Research Centre of Dijon, CHU Dijon Bourgogne, Dijon, France

Violette, Sorel, Memory Resource and Research Centre of Lille, CHRU de Lille, Lille, France

Maria, Eugenia, Soto, Memory Resource and Research Centre of Toulouse, CHU de Toulouse, Toulouse, France

Noui, Souakri, Memory Resource and Research Centre of Bordeaux, CHU de Bordeaux, Bordeaux, France

Jacqueline, Suquet, Memory Resource and Research Centre of Montpellier, CHU de Montpellier, Montpellier, France

Géraldine, Sylvestre, Memory Resource and Research Centre of Besançon, Besançon, France

Mathieu, Tafani, Memory Resource and Research Centre of Toulouse, CHU de Toulouse, Toulouse, France

Stéphanie, Taglang, Memory Resource and Research Centre of Colmar, Colmar, France

Jean-Yves, Tanguy, Memory Resource and Research Centre of Angers, CHU d'Angers, Angers, France

Lorraine, Templier, Memory Resource and Research Centre of Paris Nord, AP-HP, Paris, France

Catherine, Terrat, Memory Resource and Research Centre of Saint-Etienne, CHU de Saint-Etienne, Saint-Etienne, France

Jamila, Thabet, Memory Clinic of Avicenne, Hôpital Avicenne, Bobigny, France

Claire, Thalamas, Memory Resource and Research Centre of Toulouse, CHU de Toulouse, Toulouse, France

Nathalie, Thiery, Coordinating Centre, CIC-1401 Clinical Epidemiology, France

François, Tison, Memory Resource and Research Centre of Bordeaux, CHU de Bordeaux, Bordeaux, France

Hélène, Ton, Van, Memory Clinic of Avicenne, Hôpital Avicenne, Bobigny, France

Lucie, Toulemonde, Memory Resource and Research Centre of Marseille, CHU de Marseille, Marseille, France

Virginie, Tourbier, Memory Resource and Research of Amiens, CHU Amiens Picardie, Amiens, France

Bertrand, Toussaint, Memory Resource and Research Centre of Grenoble, CHU de Grenoble Alpes, Grenoble, France

Eve, Tramoni, Memory Resource and Research Centre of Marseille, CHU de Marseille, Marseille, France

Candice, Trocmé, Memory Resource and Research Centre of Grenoble, CHU de Grenoble Alpes, Grenoble, France

Irène, Troprès, Memory Resource and Research Centre of Grenoble, CHU de Grenoble Alpes, Grenoble, France

Anne-Cécile, Troussière, Memory Resource and Research Centre of Lille, CHRU de Lille, Lille, France

Anne, Turazzi, Memory Resource and Research Centre of Nice, CHU de Nice, Nice, France

Renata, Ursu, Memory Clinic of Avicenne, Hôpital Avicenne, Bobigny, France

Emilie, Vaillant, Memory Resource and Research Centre of Nice, CHU de Nice, Nice, France

Nathalie, Vayssière, Memory Resource and Research Centre of Toulouse, CHU de Toulouse, Toulouse, France

Pierre, Vera, Memory Resource and Research Centre of Rouen, CHU de Rouen, Rouen, France

Olivier, Vercruysse, Memory Resource and Research Centre of Lille, CHRU de Lille, Lille, France

Antoine, Verger, Memory Resource and Research Centre of Nancy, CHU de Nancy, Nancy, France

Maximilien, Vermandel, Memory Resource and Research Centre of Lille, CHRU de Lille, Lille, France

Philippe, Viau, Memory Resource and Research Centre of Nice, CHU de Nice, Nice, France

Marie-Neige, Videau, Memory Resource and Research Centre of Bordeaux, CHU de Bordeaux, Bordeaux, France

Jean-Louis, Vincent, Memory Resource and Research Centre of Lille, CHRU de Lille, Lille, France

Vincent, Visneux, Memory Resource and Research Centre of Saint-Etienne, CHU de Saint-Etienne, Saint-Etienne, France

Isabelle, Vivier, Memory Resource and Research Centre of Bordeaux, CHU de Bordeaux, Bordeaux, France

Christelle, Vlaemynck, Memory Resource and Research Centre of Clermont-Ferrand, CHU de Clermont-Ferrand, Clermont-Ferrand, France

Natacha, Vogt, Memory Resource and Research Centre of Strasbourg, CHRU de Strasbourg, Strasbourg, France

Thierry, Voisin, Memory Resource and Research Centre of Toulouse, CHU de Toulouse, Toulouse, France

Elodie, Vulliez, Memory Resource and Research Centre of Lyon, Hospices Civils de Lyon, Lyon, France

Nathalie, Wagemann, Memory Resource and Research Centre of Nantes, CHU de Nantes, Nantes, France

Caroline, Wagner, Memory Resource and Research Centre of Strasbourg, CHRU de Strasbourg, Strasbourg, France

Aziza, Waissi-Sediq, Memory Resource and Research Centre of Lyon, Hospices Civils de Lyon, Lyon, France

Sandrine, Wannepain, Memory Resource and Research of Amiens, CHU Amiens Picardie, Amiens, France

Marie-Joséphine, Waryn, Memory Clinic of Avicenne, Hôpital Avicenne, Bobigny, France

Brigitte, Weidmann, Memory Resource and Research Centre of Colmar, Colmar, France

Emilie, Wenish, Memory Resource and Research Centre of Marseille, CHU de Marseille, Marseille, France

Léocadie, Werle, Memory Resource and Research Centre of Strasbourg, CHRU de Strasbourg, Strasbourg, France

Gabrielle, Woehrel, Memory Resource and Research Centre of Strasbourg, CHRU de Strasbourg, Strasbourg, France

Jing, Xie, Memory Resource and Research Centre of Lyon, Hospices Civils de Lyon, Lyon, France

Nathanaëlle, Yeni, Laboratoire d'Imagerie Biomédicale, Sorbonne Universités, UPMC Univ Paris 06, Inserm U 1146, CNRS UMR 7371, F-75006 Paris, France NeuroSpin, I2BM, Commissariat à l'Energie Atomique, France

Michel, Zanca, Memory Resource and Research Centre of Montpellier, CHU de Montpellier, Montpellier, France

Rupestre, Zannou, Coordinating Centre, CIC-1401 Clinical Epidemiology, Clinical Research Associate, France

Jean, Zinszner, Memory Clinic of Avicenne, Hôpital Avicenne, Bobigny, France